\documentstyle[11pt,aaspp4,flushrt]{article}
\input epsf
\def\gtsima{$\, \buildrel > \over \sim \,$}
\def\ltsima{$\, \buildrel < \over \sim \,$}
\def\simgt{\lower.5ex\hbox{\gtsima}}
\def\simlt{\lower.5ex\hbox{\ltsima}}
\def\onehalf{\frac{1}{2}}
\def\sm{$\sim\,$}
\def\degs{\ifmmode^\circ\else$^\circ$\fi}

\def\msol{\ifmmode M_\odot\else$M_\odot$\fi}

\def\zhel{\ifmmode z_\odot\else$z_\odot$\fi}
\def\kms{\ifmmode{\rm km}\,{\rm s}^{-1}\else km$\,$s$^{-1}$\fi}
\def\kmsmpc{\ifmmode{\rm km}\,{\rm s}^{-1}\,{\rm Mpc}^{-1}\else km$\,$s$^{-1}\,$Mpc$^{-1}$\fi}
\def\etal{{\sl et al.}}
\def\apriori{{\em a priori}}

\def\h1{$h^{-1}$}
\def\dnsigma{$D_n$-$\sigma$}
\font\tensm=cmcsc10
\def\hii{H\kern 2.0pt{\tensm ii}}	
\def\hi{\ifmmode{\rm H\kern 2.0pt{\tensm I}}\else H\kern 2.0pt{\tensm I}\fi}	
\def\explnrD{\exp\!\left(-\frac{\left[\ln r/d \right]^2}{2 \Delta^2} \right)}
\begin{document}
\centerline{{\small To appear in {\bf ``Formation of Structure in the Universe,''}}}
\centerline{{\small A.\
Dekel and J.\ Ostriker, Eds.\ (Cambridge University Press)}}
\bigskip

\title{Measurement of Galaxy Distances}

\author{Jeffrey A. WILLICK\\
Department of Physics,
Stanford University}

\begin{abstract}
Six of the principal galaxy distance indicators are discussed:
Cepheid variables, the Tully-Fisher relation,
the \dnsigma\ relation, Surface Brightness
Fluctuations, 
Brightest Cluster Galaxies, and Type Ia Supernovae.
The role they play in peculiar velocity
surveys and Hubble constant determination is
emphasized. Past, present, and future efforts
at constructing catalogs of redshift-independent
distances are described. 
The chapter concludes with a qualitative
overview of Malmquist and related biases.
\end{abstract}

\section{Introduction}
\label{sec:intro}
The measurement of galaxy distances is one of the most fundamental
problems in astronomy. To begin with, we would simply
like to know the scale of the cosmos; we do so by determining
the distances to galaxies. Beyond this, galaxy distances are the key
to measuring the Hubble constant $H_0,$ perhaps the
most important piece of information for testing the validity of the Big Bang
model. Finally, galaxy distances are necessary if we are to study
the large-scale peculiar velocity field.  Peculiar velocity analysis is
among the most promising techniques for confirming
the gravitational instability paradigm for the origin of large-scale
structure, deducing the relative distributions of luminous and dark
matter, and constraining the value of the cosmological
density parameter $\Omega_0.$ In this Chapter, I will describe
a number of the methods used for measuring
galaxy distances, and discuss their application to
the $H_0$ and peculiar velocity problems.
When appropriate, I will comment on their
relevance to determination of other cosmological parameters
as well. The goal of this Chapter is not to present
an exhaustive review of galaxy distance measurements, 
but rather to provide a summary of where matters
stand, and an indication of what the 
next few years may bring.

\subsection{Peculiar Velocities versus $H_0$}
What it means to ``measure a galaxy's distance'' depends
on whether one is interested in studying peculiar
velocities or determining the value of the Hubble constant.
A galaxy's peculiar velocity may be estimated given its ``distance''
in \kms---the part of its radial velocity due solely
to the Hubble expansion. The same object provides an
estimate of $H_0$ only if one can measure its distance
in metric units such as megaparsecs. What this means in practice is that accurate peculiar velocity studies may be carried
out {\em today\/}, despite the fact that $H_0$ remains undetermined at
the \sm 20\% level.

Another basic distinction between velocity analysis and
the search for $H_0$ concerns the distance regimes in which they
are optimally conducted. Peculiar velocity surveys are best
carried out in the  ``nearby'' universe, where
peculiar velocity errors 
are comparable to or less than the peculiar velocities themselves. 
The characteristic amplitude of the
radial peculiar velocity, $v_p,$ is a few hundred \kms\
at all distances, whereas the errors we make in estimating $v_p$
grow linearly with distance (\S\ref{sec:TF}). 
It turns out that the ``break-even'' point occurs
at distances of \sm 5000 \kms. Although we may hope to glean 
some important information (such as bulk flow amplitudes) on
larger scales,
our ability to construct an accurate picture
of the velocity field is restricted to the region within
about 50\h1 Mpc.
In the Hubble constant problem, by contrast,
peculiar velocities are basically a nuisance. We would like them to
be a small fraction of the expansion velocity, so that
we incur as small as possible an error by neglecting them.
This is best achieved by 
using comparatively {\em distant\/}
objects, $d \simgt\ 7000\ \kms,$ as tracers of the expansion.  

On the other hand, to obtain the absolute distances 
needed to measure $H_0,$ we must first calibrate our distance
indicators {\em locally\/} ($\simlt 2000\ \kms$). This is because
the distance indicators capable of reaching the ``far field'' ($\simgt
7000\ \kms$) of the Hubble flow generally have no \apriori\
absolute calibration (cf.\ \S\ref{sec:DIs}). 
The only reliable distance indicator that
can bridge the gap between the Milky Way and the handful of
Local Group galaxies whose absolute distances are well-known,
and galaxies beyond a few Mpc,
is the Cepheid variable method (\S\ref{sec:cepheid}), which
is limited to distances $\simlt 2000\ \kms.$  As
a result, {\em Hubble
constant measurement is inherently a two-step process:} local
calibration in galaxies with Cepheid distances,
followed by distance measurements in the far field where
the effect of peculiar velocities is small.
The local calibration
step is unnecessary in peculiar velocity studies.

Although peculiar velocity surveys 
and $H_0$ measurement differ in the ways just
discussed, the two problems are, ultimately,
closely related.
Many distance indicator methods have been
and are being used for both purposes. Indeed, a
distance indicator calibrated
in \kms\ may be
turned into a tool for measuring $H_0$
simply by knowing the distances in Mpc to a few well-studied objects
to which it has been applied.
This Chapter will thus be organized not around the peculiar
velocity-$H_0$ distinction, but rather around methods
of distance estimation.  

\subsection{Distance Indicators}
\label{sec:DIs}
Measuring the distance to a galaxy almost always involves
one of the following properties of the propagation
of light: (1) The apparent brightness of a source falls of inversely
with the square of its distance; (2) The angular
size of a source falls off inversely with its distance.
As a result, we can determine the
distance to an object by knowing its intrinsic
luminosity or linear size, and then comparing 
with its apparent brightness
or angular size, respectively.
If all objects of
a given class 
had approximately the same absolute magnitude, we could
immediately determine their distances simply by comparing
with their apparent magnitudes. Such objects are
called {\em standard candles.} Similarly, classes of objects whose
intrinsic linear sizes are all about the same are known as ``standard
rulers.'' True
standard candles or rulers are, however, extremely rare
in astronomy. It is much more often the case that
the objects in question possess another, {\em distance-independent\/}
property from which we infer their absolute magnitudes
or diameters. For example, the rotation velocities of spirals
galaxies are good predictors of their luminosities (\S\ref{sec:TF}),
while the central velocity dispersions and surface brightnesses
of ellipticals together are good predictors of their diameters (\S\ref{sec:dnsigma}).
Whether standard candles or rulers,
or members of the more common second category, objects whose
absolute magnitudes or diameters we can somehow ascertain
are known as {\em Distance Indicators,} or DIs.

Absolute calibration of most DIs is not
straightforward. 
One discovers
that a particular distance-independent
property is a good predictor of absolute magnitude because
it is well correlated with the {\em apparent}
magnitudes of objects lying at a common distance---in
a rich cluster of galaxies, for example.
Such data may be used to determine the mathematical
form of the correlation (e.g., linear with a given slope).
However, the cluster distance in most cases is
not accurately known. Thus,
the predicted absolute magnitude corresponding
to a given value of the distance-independent
property---the ``zero point'' of the DI---remains undetermined up to a constant,
assuming one has no rigorous, \apriori\ physical theory
of the correlation, as is usually the case (but see below).
Any distances obtained from the DI at this point
will be in error by a fixed scale factor.
This situation is obviously unacceptable
for the Hubble constant problem, in which absolute
distances are required. The remedy is to determine
the zero point of the DI by applying it to
galaxies whose true distances have been
determined by an independent technique 
(e.g., Cepheid variables), as discussed above.
Such a DI is said to be ``empirically'' calibrated.

For peculiar velocity surveys, the situation is simpler because
absolute calibration is not required. However,
the DI must still be calibrated such that it yields distances
in \kms, the radial velocity due to Hubble flow. For this,
one must apply the DI to many
galaxies, widely enough distributed around the sky and
at large enough distances that peculiar velocities tend to
cancel out. Only then can redshift be taken
as a good indicator on average of distance in \kms,
and a calibration in velocity units thereby obtained
(Willick \etal\ 1995, 1996).
Empirical DI calibration, in this sense, is needed even for peculiar velocity work.

DIs of this sort tend to make some people
nervous. They argue that a good distance
estimation method should be based on solid, calculable
physics. There are, in fact, a few such techniques. 
One involves
exploitation of the Sunyaev-Zeldovich effect in clusters,
in which comparison of Cosmic Microwave Background distortions
and the X-ray emission produced by hot, intracluster
gas yields the physical size of the cluster (cf.\
Rephaeli 1995 for a comprehensive review).
Another method involves modeling time delays
between multiple images of gravitationally lensed background
objects (see the Chapter by Narayan and Bartelmann in this
volume).  
Other DIs for which theoretical absolute calibration may be possible
are Type II Supernovae, whose expansion
velocities may be related to luminosities 
(Montes \& Wagoner 1995; Eastman, Schmidt, \& Kirshner 1996), and Type Ia Supernovae,
whose luminosities may be calculated
from theoretical modeling of the explosion mechanism (Fisher, Branch,
\& Nugent 1993).
Such approaches are indeed promising, and will undoubtedly
contribute to the measurement of $H_0$ over the next decade.
However, at present these methods should  be considered preliminary.
Some of the underlying physics remains to be worked out, and many of the
underlying assumptions will need to be tested. Furthermore,
the data needed to implement such techniques are currently rather scarce. 
With the exception
of Type Ia Supernovae (discussed in \S\ref{sec:SN} in their traditional,
empirical context), I will not discuss these methods further in this Chapter.

I will focus instead on methods that require empirical 
calibration. These DIs arise from astrophysical correlations
Nature was kind enough to provide us with, but mischievous
enough to deny us a full understanding of.  The canonical
wisdom, which states that we need hard physical theory
that explains a DI in order to trust it, is a bit
too exacting given our present theoretical and observational capabilities.
We should conditionally trust our empirical DIs while recognizing the
uncertainties involved.
In particular, we must 
remember that since they possess no \apriori\
absolute calibration, they must
(for measuring $H_0$) be carefully calibrated locally.
We must also remain open to the possibility that they
may not behave identically in different
environments and at different redshifts.
Our belief in their utility should be tempered by a healthy
skepticism about their universality, and the distance
estimates we make with them subjected to continuing 
consistency checks.

\section{Cepheid Variables}
\label{sec:cepheid}
Cepheids variable stars have been fundamental to unlocking the cosmological
distance scale since Henrietta Leavitt used them in 1912
to estimate the distances to the Magellanic Clouds. 
Of the various DIs discussed in this Chapter, the Cepheid method is
the only one involved in the Hubble constant but not the peculiar
velocity problem. Indeed, it is probably safe to say that the {\em raison
d'\^etre\/} for Cepheid observations is the ultimate determination of $H_0.$
They will do so, however, in conjunction with, not independently of,
the secondary distance indicators discussed in later sections.

\begin{figure*}
\vspace{0.3cm}
\centerline{\epsfxsize=4.0 in \epsfbox{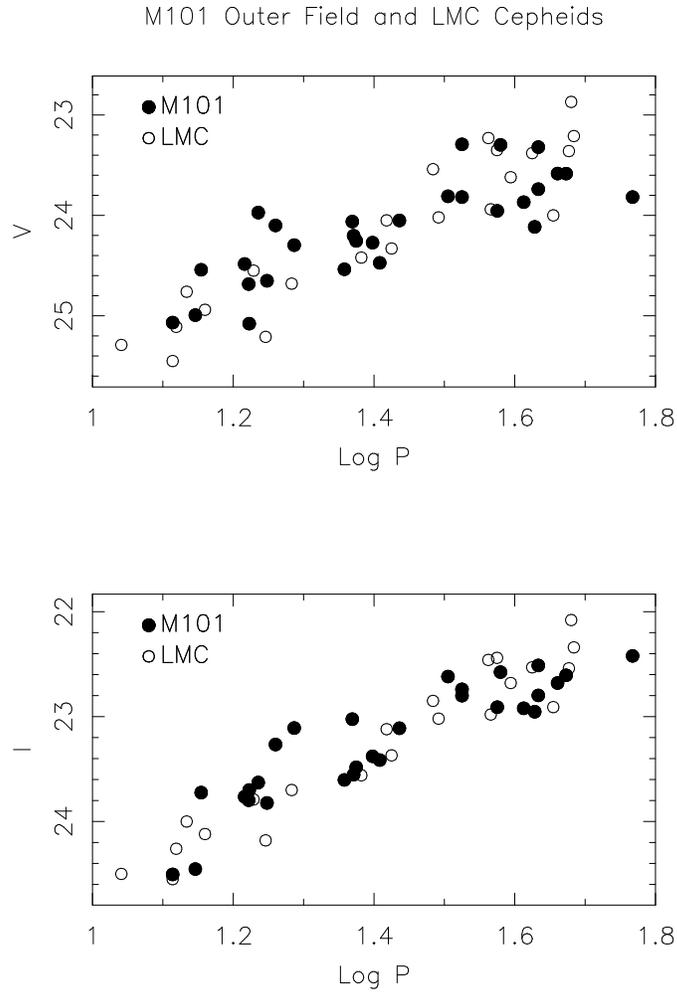}}
\label{fig:cepheids}
\caption{Cepheid variable Period-Luminosity (PL) relations
for the $V$ and $I$ bandpasses. Data for 
M101 and the Large Magellanic Cloud are shown.
Adapted from Ferrarese \etal\ (1996).}
\end{figure*}

Cepheids are post-main sequence stars that occupy the instability
strip in the H-R diagram. They pulsate according
to a characteristic ``sawtooth'' pattern, with periods that
can range from a few days to a good fraction of a year.
Cepheids exhibit an excellent
correlation between  
mean luminosity (averaged over a pulsation cycle) and
pulsation period. This
correlation is shown in Figure~\ref{fig:cepheids} for Cepheids recently
measured by the Hubble Space Telescope (HST)
in the nearby galaxy M101 (solid points), and also for
Cepheids in the Large Magellanic Cloud (LMC) as they would
appear if the LMC lay at the distance of M101. It is apparent
that the correlation is extremely similar for the two galaxies. 
Modern calibrations of the Cepheid {\em Period-Luminosity\/} (P-L)
relation in the $V$ and $I$ bandpasses are 
\begin{equation}
M_V = -2.76\left[\log(P)-1.0\right]-4.16 
\label{eq:cephV}
\end{equation}
and
\begin{equation}
M_I = -3.06\left[\log(P)-1.0\right]-4.87
\label{eq:cephI}
\end{equation}
(Ferrarese \etal\ 1996).
The absolute zero points of these P-L relations have been obtained
by observing Cepheids in the
the Large and Small Magellanic Clouds, whose distances
are known from main sequence fitting (Kennicut, Freedman, \& Mould 1995). 
Equations~(\ref{eq:cephV}) and~(\ref{eq:cephI}) show that Cepheid
variables {\em are intrinsically bright stars.}
Even short-period ($P \simeq 10^d$) Cepheids have absolute
magnitudes $M_V<-4,$ and long-period ($P \simeq 50$--$100^d$)
Cepheids are 2--3 magnitudes brighter still.
It follows that individual Cepheid stars can be observed at relatively
large distances. Indeed, with the HST
Cepheids can be observed out to the distance of the Virgo
cluster and possibly beyond. To be useful as distance
indicators, however, Cepheids cannot be merely
{\em detected.} 
Because they are found
in crowded fields, they must be well above the limit of
detectability at all phases in order to be accurately
photometered. 
These stringent requirements
place a limit of $m_V\simeq 26$ mag, much brighter than the HST
detection limit of \sm 30 mag, for distance scale work using Cepheids.

Cepheids yield distances to their host galaxies by comparison of
their absolute magnitudes, inferred from the P-L relation,
with their observed apparent magnitudes.
Specifically, 
the distance to the host galaxy is obtained by
fitting equations~(\ref{eq:cephV}) and~(\ref{eq:cephI}),
plus a distance modulus offset $\mu=5\log(d/10)$ (where
$d$ is in parsecs), to the 
observed $m_V$ and $m_I$ versus $\log(P)$ diagram. (The
same exercise may of course be carried out 
in other bandpasses as well.) An important advance
has been made in recent years by Freedman, Madore, and
coworkers, who have developed a method for correcting
for extinction in the host galaxies (Freedman \& Madore 1990;
Freedman, Wilson, \& Madore 1991). 
In brief, the photometry is
done in several bandpasses, and the magnitudes corrected
for an assumed value of the extinction within the host galaxy.
The distance modulus is determined for each
bandpass, as described above.
The value of extinction which brings
the distance moduli in the various bands into
agreement is assumed to be the correct one. This technique
works best when data for a wide range of
wavelengths, including if possible the near infrared, are available.

The great utility of Cepheids has been recognized
in the designation of an HST Key Project to measure
Cepheid distances for  20 nearby galaxies. This program, led by
Wendy Freedman, Robert Kennicut, and Jeremy Mould, 
produced its first results in late 1994. As of this
writing (July 1996), Cepheid distances from the Key Project 
are available for only a handful of galaxies. Distances for
the remaining galaxies are expected to become available over
the next few years. The results that have received
the greatest attention to date involve the
Virgo cluster galaxy M100, in which
over 50 Cepheid variables have now been accurately
measured (Freedman \etal\ 1994;
Mould \etal\ 1995; Ferrarese \etal\ 1996). 
Fitting the universal P-L relations above
to the M100 data yields a distance of
$16.1\pm 1.3$ Mpc. When combined with a suite of
assumptions concerning the morphology and peculiar
velocity of the Virgo cluster, this distance suggests
a Hubble constant of about 85 \kmsmpc (Freedman \etal\ 1994).

Unfortunately, the Hubble constant estimate obtained from M100 has received
undue attention. This is understandable, given that determination
of $H_0$ is the long-term aim of the Key Project. And, of course,
values of $H_0$  in excess of \sm 75 \kmsmpc\
are difficult to square with most estimates of
the age of the universe based on its oldest constituents.
But as the Key Project group has emphasized (Kennicutt, Freedman,
\& Mould 1995), a single
galaxy in the Virgo cluster with a good Cepheid distance
does not allow one to estimate the Hubble constant with
any accuracy. In fact, the Virgo cluster is a 
poor laboratory in which to estimate $H_0$ no matter
how many galaxies one has Cepheid distances for. The
reasons are simple: Virgo's depth is a good fraction (\sm 30\%)
of its distance, and its peculiar velocity is likely to
be a good fraction (\sm 20--30\%) of its Hubble velocity.
The velocity/distance ratio of any single Virgo object,
or even group of objects, may therefore be a poor approximation
of $H_0,$ and it is difficult to gauge the systematic errors that affect it.

Thus, Cepheid variables will not themselves be
used to measure $H_0.$ Instead, they will be used to obtain
accurate distances for several tens of galaxies within about 20\h1 Mpc.
These galaxies will in turn
serve as calibrators
for the {\em secondary distance indicators,} such as Type 1a Supernovae
and the Tully-Fisher relation, that are applicable in
the far field of the Hubble flow (and occupy the
remainder of this Chapter). Initial steps
in this direction have already been taken by Sandage, Tammann,
and coworkers (Sandage \etal\ 1996), who used HST Cepheid
distances (their own, not those of the Key Project) to calibrate
historical and contemporary Type Ia Supernovae. When they apply this calibration
to distant Type Ia SNe (Tammann \& Sandage 1995), 
they derive $H_0=56$--58
\kmsmpc\ (the lower value applies to $B$-band,
and the higher value to $V$-band, measurements; Sandage
\etal\ 1996).
There is considerable
controversy, however, surrounding the calibration of the historical
photometry used in the SNe Ia calibration. Furthermore, the Sandage 
group has neglected the correlation between the peak luminosity
of SNe Ias and the width of their light curves, an effect which
now appears important (\S\ref{sec:SN}).
Until these issues are
resolved, and agreement between the Sandage and HST Key Project
groups on local Cepheid distances achieved,
estimates of $H_0$ based on this approach should be considered preliminary.

\section{The Tully-Fisher Relation for Spiral Galaxies}
\label{sec:TF}
It has been stated that the Tully-Fisher (TF) relation is the ``workhorse''
of peculiar velocity surveys. One can anticipate
a time in the not so distant future when more accurate techniques may 
supplant it, but for the next few years at least, the TF
relation is likely to remain the most widely used distance indicator
in cosmic velocity studies. Its role in such studies to date
has been, in fact, too large to be reviewed here, and interested readers
are referred to Strauss \& Willick (1995), \S 7.
Several recent developments 
are discussed in \S\ref{sec:catalogs} below.

The TF relation is one of the most fundamental properties of spiral galaxies.
It is the empirical statement of an approximately power-law 
relation between luminosity and rotation velocity. 
Specifically, it is found that
\begin{equation}
L \propto v_{{\rm rot}}^\alpha\,,
\label{eq:tflum}
\end{equation}
or, using the logarithmic formulation preferred by working astronomers,
\begin{equation}
M = A - b\eta\,.
\label{eq:tfmag}
\end{equation}
In equation~(\ref{eq:tfmag}) 
$M=-2.5\log(L)+{\rm const.}$ is the absolute magnitude, and
the {\em velocity width parameter\/}
$\eta\equiv\log(2v_{{\rm rot}})-2.5,$ where $v_{{\rm rot}}$ is expressed
in \kms, is a useful dimensionless measure of rotation velocity. 

\begin{figure*}
\centerline{\epsfxsize=4.5 in \epsfbox{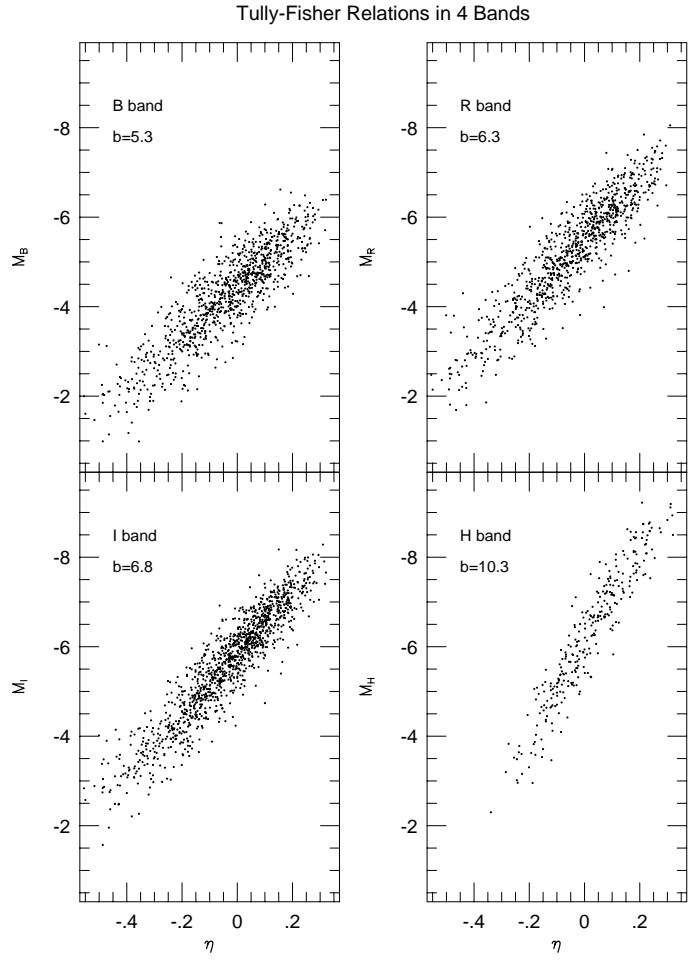}}
\label{fig:4tf}
\caption{TF relations in four bandpasses. 
The absolute magnitudes are given in units
such that $d=10^{0.2(m-M)}$ is the galaxy distance in \kms.
Adapted from
Strauss \& Willick (1995).}
\end{figure*}
An important fact, not always sufficiently appreciated,
is that the power law
exponent $\alpha$ does not have a unique value. 
The details of both the photometric and spectroscopic
measurements affect it. A typical result found in
contemporary studies is $\alpha\simeq 3.$ The corresponding value
of the ``TF slope,'' $b,$ is $\sim 7.5.$ However, slight changes
in the details of measurement can result in significant changes
in $b.$ This is illustrated in Figure~\ref{fig:4tf}, in which
TF relations in four bandpasses are plotted. The optical
bandpasses ($B,$ $R,$ and $I$) all represent data from
the sample of Mathewson \etal\ (1992). In each case the slope
is $<7.$ This is because Mathewson \etal\ defined their
velocity widths rather differently than most observers, with
the result that small velocity widths are made smaller still relative
to other width measurement systems, while large velocity widths
are unchanged in comparison with other systems. If one were
to transform the Mathewson \etal\ widths to those of standard
systems, one would obtain and $I$-band TF slope of $\sim 7.7,$
comparable to other $I$-band samples (Willick \etal\ 1997).
Note, however, that the slope increases steadily from $B$ to $I.$
This reflects a general trend of increasing TF slope toward
larger wavelengths, which was noted over a decade ago (Bottinelli
\etal\ 1983).
The $H$-band data come from the compilation of Aaronson \etal\
(1982), as reanalyzed by Tormen \& Burstein (1995) and Willick \etal\ (1996).
The $H$-band TF slope is considerably greater than its optical counterparts.
In part this reflects the trend just noted. However, to a greater
extent it is 
due to the relatively small apertures within which the $H$-band
photometry is done: the TF slope increases as the photometric
aperture size is {\em decreased\/}
(Willick 1991). Infrared TF samples in which {\em total\/} magnitudes
were measured exhibit TF slopes 
not much in excess of the optical values
(Bernstein \etal\ 1994). 

The numerical value of the TF zero point $A$ has no absolute
significance in itself, reflecting mainly the photometric
system in which the TF measurements are done.
Absolute magnitudes measured
in different bandpasses can differ numerically by a few magnitudes 
for a given object.
Clearly, this must have no meaning for distance measurements,
and it is the zero point that absorbs such differences.
For any given measurement system, however, the value of
$A$ is highly significant.  To see this,
consider how the TF relation is used to
infer distances and peculiar velocities. Given a measured apparent
magnitude $m$ and width parameter $\eta,$ one infers the
distance modulus to the galaxy as $\mu=m-(A-b\eta).$ The
corresponding distance $d\propto10^{0.2\mu}.$ It follows that
an error $\delta\! A$ in the TF zero point thus corresponds to a  
fractional distance error $f=10^{-0.2\delta\! A}.$ A Hubble constant
inferred from such distances will then be off by a factor $f^{-1}.$
For peculiar velocities, calibration of the
TF relation consists in choosing $A$ such that $d\equiv10^{0.2[m-(A-b\eta)]}$
gives a galaxy's distance in \kms. There is no requirement that it
yield the distance in Mpc. Nonetheless, as mentioned above,
zero point calibration errors are still possible. Errors in
$A$ produce distances in \kms\ that
differ by a fraction $f$ from the true Hubble velocity,
with resultant
peculiar velocity errors $\delta v_p = -fd.$

The TF relation has a rich history (cf.\ Bottinelli \etal\ 1983) which
can not be discussed in any detail here. 
Its discovery is generally
credited to Tully \& Fisher (1977), who were the first to suggest
a linear correlation between absolute magnitude and log rotation
speed. Until the late 1980s, the velocity widths used in TF studies
were generally obtained from analysis of the 21 cm line profiles,
rather than from direct measurements of rotation
curves. Thus, the
TF relation was (and to some extent remains) closely associated
with 21 cm radio astronomy. There is no inherent connection,
however, between the TF relation and the 21 cm line.
The TF relation is also closely associated
in the minds of many with infrared magnitudes. This is largely
due to the pioneering $H$-band ($1.6\,\mu$m) TF work of a group  
headed by the late Marc Aaronson in the late 1970s through the
mid-1980s (e.g., Aaronson \etal\ 1980, 1982, 1986). 
This group argued that the infrared was better than the optical
for TF purposes because infrared magnitudes
are less subject to internal and Galactic extinction 
(\S\ref{sec:TFdetails}), and 
because they are sensitive mainly to
the old stellar population that best traces mass. 
While these arguments are 
true at some level, work over the last decade has demonstrated
that CCD imaging photometry in red passbands (e.g., $R$ or $I$)
results in TF relations that, empirically, work as well or better
than the $H$-band version (Pierce \& Tully 1988, 1992; Willick 1991;
Han 1992; Courteau 1992; Bernstein 1994; Willick \etal\ 1995,1996).

\subsection{The TF Relation and Galaxy Structure}
Many workers have attempted to ``explain'' the TF relation on the
basis of physical principles and models of galaxy formation.
While these attempts can claim some modest successes, it is probably
fair to say that a true explication of the TF relation remains elusive.
One can argue heuristically that something like the TF relation
must exist: Assuming that luminosity is proportional to mass,
and that a virial relation $v^2\sim GM/R$ holds for spirals, 
it follows that $L \propto M \propto Rv^2.$ If one further notes that
spirals have characteristic surface brightnesses $I\propto L/R^2$
that varies little from galaxy to galaxy, then $R\propto L^{\onehalf},$ and
it follows that $L\propto v^4.$ This was indeed the power-law
exponent (i.e., $b\simeq 10$)
originally found by the Aaronson group, and the argument
seemed reasonable to them (Aaronson, Huchra, \& Mould 1979).

However, quite a few loose ends remain.
First, as noted above, contemporary measures of the TF slope
suggest that the exponent is closer to 3 than to 4. The aperture
and wavelength dependences noted above  
tell us that the TF slope is
not determined strictly by idealized dynamics, but depends
also on the details of the distribution---in both space
and wavelength---of the starlight emitted by the galaxy.
Furthermore, while a number of theoretical approaches
can approximately predict the TF slope, no realistic
model has successfully accounted for its rather
small (\sm 0.3 mag; see below) intrinsic scatter 
(Eisenstein \& Loeb 1996).

Another, more fundamental, problem is that the TF relation is
evidently connected with the phenomenon of flat rotation curves (RCs)
exhibited by most spiral galaxies. Were the RCs not flat, there
would be no well-defined rotation velocity, and one would expect
the TF relation to require a very specific type of velocity
width measurement. In fact, a well-defined TF relation is
found regardless of the specific algorithm for measuring
rotation velocity (although slight variations of slope
and zero point arise as a result of algorithmic
differences). Whether one measures \hi\ profile widths,
asymptotic rotation velocities, ``isophotal'' rotation velocities
(Schlegel 1996), or maximum rotation velocities, basically
similar TF relations result. Because the origin of flat rotation
curves is connected with the nature of dark
matter, it follows that we cannot fully understand the TF
relation until we understand how galaxies form in their
dark matter halos.

\subsection{Applying the TF Relation: A Few Details}
\label{sec:TFdetails}

Widely appreciated by practitioners of the TF relation,
but often hidden to the wider astronomical public, are the
careful correction procedures applied to
the magnitudes
and velocity widths that go into the TF relation. 
Probably the most important step is correction
for projection of the disk on the plane of the sky. The observed
velocity width is smaller by a factor $\sin(i),$ where
$i$ is the galaxy inclination, than the intrinsic value. Observers
correct for this by esimating $i$ from the
apparent ellipticity of the galaxy disk. Modern CCD observations
allow one to fit elliptical isophotes to the galaxy image; these
isophotes typically converge to a constant ellipticity $\varepsilon$
in the outer regions. When CCD surface photometry is not available
(as is the case for many of the older infrared data), one simply
takes $\varepsilon=1-b/a,$ where $a$ and $b$ are the major and
minor axis diameters of the galaxy obtained
from photographic data. Whichever method is used, the inclination
$i$ is taken to be a function of $\varepsilon.$ A typical formula
employed is 
\begin{equation}
\cos^2 i = \left\{ \begin{array}{ll}
\frac{(1-\varepsilon)^2-(1-\varepsilon_{max})^2}{1-(1-\varepsilon_{max})^2}\,, &
\varepsilon < \varepsilon_{max}\,; \\
0\,, & \varepsilon \geq \varepsilon_{max}\,,
                   \end{array} \right.
\label{eq:incleps}
\end{equation}  
where $\varepsilon_{max} \simeq 0.8$ is the ellipticity
exhibited by an edge-on spiral. It is apparent that formulae
such as equation~(\ref{eq:incleps}) are at best approximations,
hopefully valid in a statistical sense. However, they are usually
the best we can do, and are certainly far better than doing
nothing. Still, the inclination
correction to the widths can go seriously awry at small
inclinations, and most TF samples exclude galaxies with $i\simlt 40\degs.$ 

\def\cint{C_{{\rm int}}}
Another tricky detail of the TF relation is correcting
for {\em internal extinction.} As a spiral galaxy tilts
toward edge-on orientation, it becomes 
fainter. Since spirals are viewed at a range of orientations,
it is important to correct for this effect. The most
widely used correction is to brighten the
raw magnitudes by an amount $\cint\times\log(a/b),$ where $\cint$
is the {\em internal extinction coefficient.}
Studies have shown that $\cint$ is bandpass-dependent,
as one might expect. However, in the optical red
($R$ and $I$ bandpasses), the wavelength-dependence
is very weak, and $\cint\simeq 1$ is a good approximation
(Burstein, Willick, \& Courteau 1995; Willick \etal\ 1996,1997).
A controversial question is whether internal extinction
depends on any galaxian property other than
axial ratio. Giovanelli \etal\ (1995)  argued that it
is luminosity-dependent, but Willick \etal\ (1996) reached
the opposite conclusion through a TF-residual analysis.
This issue merits further consideration
in the future.

\subsection{The TF Scatter}

\def\sigtf{\sigma_{{\rm TF}}}
Of great importance to applications of the TF relation 
is its scatter $\sigtf,$
the rms magnitude dispersion about the mean relation
$M(\eta).$ This scatter is composed of three basic
contributions: magnitude and velocity width measurement
errors, and intrinsic or ``cosmic'' scatter. 
Of the three, recent analyses have suggested that
the second and third are about equally important,
contributing $\sim 0.25$--$0.30$ mag each (Willick \etal\ 1996).
Photometric measurement errors are quite small
in comparison. Thus, the overall TF scatter is
about 0.4 mag. It is significant that $\sigtf$
determines not only random
distance errors ($\frac{\delta d}{d} \simeq 0.46\,\sigtf$),
but also systematic errors associated with
statistical bias effects (\S\ref{sec:bias}). Knowing $\sigtf$
is therefore crucial for assessing the reliability
of TF studies. (An analogous statement applies to the
scatter of the other DIs discussed in this Chapter as well.)

I would be remiss if I did not mention that the TF scatter
remains controversial. Estimates of $\sigtf$ have varied
widely in the last decade. Bothun \& Mould (1987)
suggested that $\sigtf$ could be made
as small as $\simlt 0.25$ mag with a velocity width-dependent
choice of photometric aperture. Pierce \& Tully (1988)
also found $\sigtf\simeq 0.25$ using CCD data in the 
Virgo and Ursa Major clusters. Willick (1991) and Courteau
(1992) found somewhat higher but still small values of
the TF scatter ($\sigtf = 0.30$--0.35 mag). Bernstein \etal\ (1994)
found the astonishing value of 0.1 mag for the Coma Cluster
TF relation using $I$-band CCD magnitudes and carefully measured
\hi\ velocity widths.

Unfortunately, these relatively low values have
not been borne out by later studies using more complete samples.
Willick \etal\ (1995,1996,1997) calibrated TF relations for
six separate samples comprising nearly 3000 spiral galaxies,
and found typical values of $\sigtf\simeq 0.4$ mag for the
CCD samples. Willick \etal\ (1996) argued that the large sample
size and a relatively conservative approach to excluding outliers
drove up earlier, optimistically low estimates of the TF scatter.
Other workers, notably Sandage and collaborators (e.g., Sandage 1994;
Federspiel \etal\ 1994) have taken an even more pessimistic
view of the accuracy of the TF relation, suggesting that
typical spirals scatter about the TF expectation by
0.6--0.7 mag.

How can one reconcile this wide range of values? 
At least part of the answer lies in different workers' 
preconceptions and preferences. Those excited at the possiblity
of finding a more accurate way of estimating distances tend
to find low ($\sigtf \simlt 0.3$ mag) values. Those who doubt
the credibility of TF distances tend to find high ($\sigtf \simgt 0.5$ mag)
ones. It is possible to arrive at such discrepant results in part
because the samples differ so dramatically. Perhaps it is
only justified to speak of a particular value of the TF scatter for a given set of
sample selection criteria; hopefully, this issue will be
clarified in the years to come.

There is one galaxian property with which the TF scatter
demonstrably appears to vary, however, and that is luminosity (velocity
width). Brighter galaxies exhibit a smaller TF scatter than
fainter ones (Federspiel \etal\ 1994; Freudling \etal\ 1995;
Willick \etal\ 1997). Part of this effect is undoubtedly
due to the fact that the errors in $\eta=\log\Delta v -2.5$
go as $(\Delta v)^{-1},$ if errors in $\Delta v$ itself
are roughly constant as is most likely the case. Such velocity width
errors translate directly into a TF scatter that increases with
decreasing luminosity. A careful study of whether the
{\em intrinsic\/} TF scatter varies with luminosity has not
yet been carried out.

\subsection{Future Directions}
\begin{figure*}
\centerline{\epsfxsize=6.0 in \epsfbox{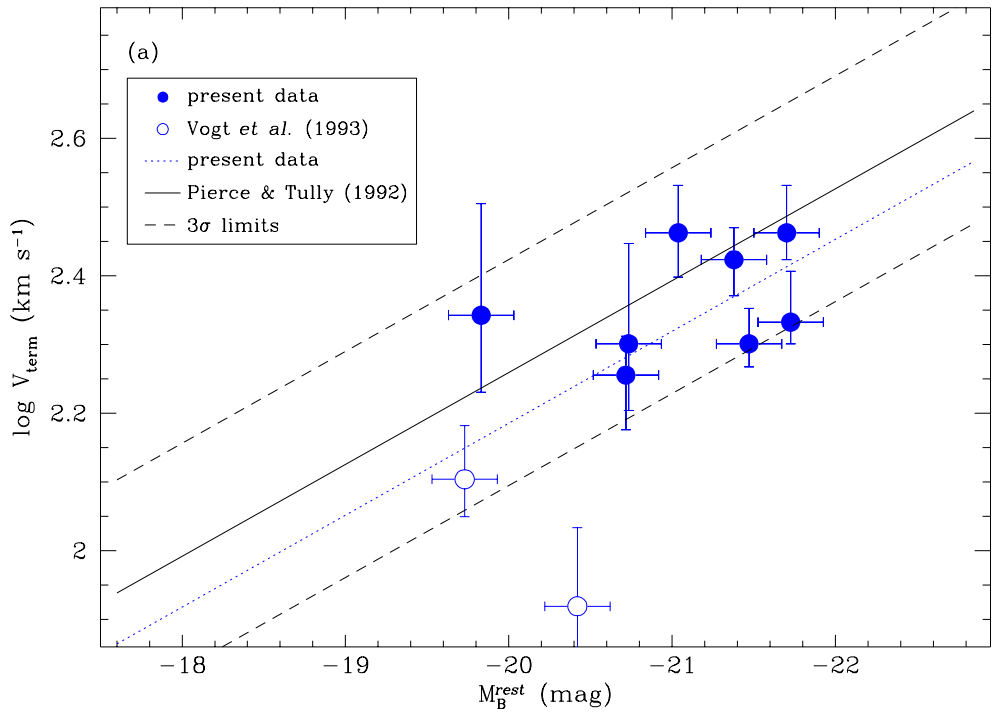}}
\label{fig:vogt}
\caption{Rotation velocity versus absolute magnitude for
spiral galaxies at a median redshift of \sm 0.5. This
figure has been adapted from Vogt \etal\ (1996).}
\end{figure*}

An intriguing recent development
been application of the TF relation to relatively
high-redshift galaxies. This has been made possible by
the advent of large-aperture telescopes 
capable of measuring rotation curves
out to redshifts of $z\simeq 1.$ Vogt \etal\ (1996) 
measured rotation curves and magnitudes for nine
field galaxies in the redshift range $0.1 \simlt z \simlt 1$
using the Keck 10-meter telescope. They found 
such objects obey a TF relation similar to that of
local objects, with only a modest shift ($\Delta M_B\simlt 0.6$ mag)
toward brighter magnitudes. This is illustrated in Figure~\ref{fig:vogt},
in which the Vogt \etal\ data are plotted along with the
TF relation derived by Pierce \& Tully (1992). However, a very
different conclusion has been reached by Rix \etal\ (1996), who
combined photometry with fiber-optic spectroscopy of spirals
at moderate ($z\simeq 0.25$) redshift. Rix \etal\ conclude that
even at such modest look-back times, spiral galaxies are
significantly ($\sim 1.5$ mag) brighter than their local
counterparts. If the TF relation is to be applied to problems such
as peculiar velocities at high redshift or estimation of $q_0,$
its evolution with redshift will have to be
understood. This is an observatonal problem which deserves,
and will undoubtedly receive,
considerably more attention in the near future. 

\section{The \dnsigma\ and Fundamental Plane Relations for Elliptical Galaxies}
\label{sec:dnsigma}
If the TF relation has been the workhorse of modern
velocity field studies, the \dnsigma\ relation has
been a short step behind. The closest analogue to
the TF relation for elliptical galaxies is actually
the predecessor of \dnsigma, the Faber-Jackson (FJ) relation.
FJ expresses the power-law correlation between an elliptical
galaxy's luminosity and its internal velocity dispersion,
\begin{equation}
L \propto \sigma_e^\alpha\,,
\label{eq:fj}
\end{equation}
where the exponent $\alpha$ was found empirically to
be $\sim 4\pm 1$ (Faber \& Jackson 1976; Schechter 1980; 
Tonry \& Davis 1981). Although discovered around the
same time, and viewed as closely related in physical origin,
TF and FJ were not considered equivalently good distance
indicators. It was clear from the outset that the scatter
in the FJ relation was about twice that of the TF relation,
on the order of 0.8 mag. Thus, while the TF relation flourished
in the early 1980s as a tool of distance measurement (\S\ref{sec:TF}), 
elliptical galaxy surveys focused more on
the stuctural and dynamical implications of the FJ relation.

These surveys bore unexpected fruit, however, in the latter
part of the 1980s. Two groups conducting 
surveys of ellipticals arrived independently at a new
result: the FJ correlation could be considerably tightened
by the addition of a third parameter, namely, surface brightness
(Djorgovski \& Davis 1987; Dressler \etal\ 1987).
In its modern incarnation, the new correlation has become
known as the \dnsigma\ relation: a power-law correlation
between the {\em luminous diameter\/}
$D_n$ and the internal velocity dispersion $\sigma,$
\begin{equation}
D_n \propto \sigma_e^\gamma\,,
\label{eq:dnsigma}
\end{equation}
where $\gamma = 1.20\pm 0.10$ (Lynden-Bell \etal\ 1988). ($D_n$ is
defined as the diameter within which the galaxy has a given
mean surface brightness. As such, it implicitly incorporates the third
parameter into the correlation.)

\begin{figure*}
\vspace{1cm}
\centerline{\epsfxsize=4.0 in \epsfbox{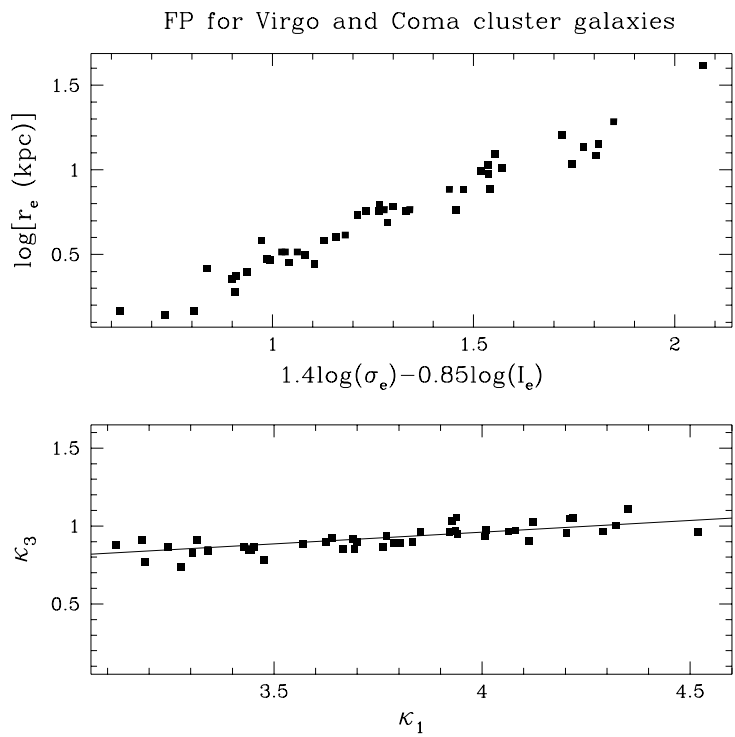}}
\caption{Two versions of the Fundamental Plane for the Virgo
and Coma ellipticals studied by Bender, Burstein, \& Faber (1992).
Further details are given in the main text. (The data used for
these figures were kindly provided by D.\ Burstein.)} 
\label{fig:FP}
\end{figure*}

More broadly, \dnsigma\ and its variants may be viewed
as manifestations of the {\em Fundamental Plane (FP) of Elliptical
Galaxies,} a planar region in the three-dimensional space 
of structural parameters
in which normal ellipticals are found.
One expression of the FP relates effective diameter
to internal velocity dispersion and central
surface brightness,
\begin{equation}
R_e \propto \sigma_e^\alpha I_e^{-\beta}\,.
\label{eq:FP}
\end{equation}
An early determination of the parameters $\alpha$ and $\beta$ using
$B$-band photometry gave
$\alpha\simeq 1.4,$ $\beta\simeq 0.9$ (Faber \etal\ 1987). 
More recently,
Bender, Burstein, \& Faber (1992) found $\alpha=1.4,$ $\beta=0.85$
using $B$-band data for a sample of Virgo and Coma cluster ellipticals; 
the upper panel of Figure~\ref{fig:FP}
shows the FP for this sample.
A recent $R$-band FP analysis by
the EFAR group (Wegner \etal\ 1996)
is $\alpha = 1.23,$ $\beta=0.72.$
A measurement based on Gunn $r$-band photometry
(Jorgensen, Franx, \& Kjaergaard 1996)
yields a similar value of $\alpha$ $(1.24\pm 0.07)$ but a somewhat
different value of $\beta$ ($0.82\pm 0.02$), perhaps due
to the slightly different bandpass used.
Pahre, Djorgovski, \& de Carvalho (1995) have recently
carried out the first analysis of the FP using $K$-band
photometry, finding $\alpha=1.44\pm 0.04,$ $\beta=0.79\pm 0.04.$ 

The two-dimensionality of the loci in parameter
space occupied by ellipticals
actually makes the FP relations, including \dnsigma, somewhat
less mysterious than the one-dimensional TF sequence. As noted
by Faber \etal\  (1987), such two-dimensionality is expected on
virial equlibrium grounds alone. Unlike the TF relation, therefore,
the FP is
not obviously related to the relative distribution of 
luminous and dark matter.
If the virial theorem were truly {\em all\/} there
was to the FP, however, one would find
$r_e \propto \sigma_e^2 I_e^{-1}.$ The fact that
the FP coefficients differ significantly from these
values implies that the mass-to-light ($M/L$) ratios
of ellipticals vary slowly as a function of mass. In particular,
the observed FP relations indicate that 
\begin{equation}
(M/L) \propto (M)^\epsilon \, ,
\label{eq:FPM}
\end{equation}
with $\epsilon \simeq\,$0.15--0.20.
Bender, Burstein \& Faber (1992)
have used this fact to look at the FP in a different way. They
define coordinates $(\kappa_1,\kappa_2,\kappa_3),$ each of which is
a normalized linear combination of $\log(\sigma_e),$ $\log(r_e),$
and $\log(I_e).$ The definitions are such that $\kappa_1$ is
proportional to $\log(M)$ and $\kappa_3$ is proportional to $\log(M/L).$ 
A plot of $\kappa_3$ versus $\kappa_1,$ shown in the bottom panel
of Figure~\ref{fig:FP}, is very nearly an edge-on projection
of the FP. The line drawn through the data is $\kappa_3 = 0.15\kappa_1 +0.36.$
A $\kappa$-space analysis of the properties
of elliptical galaxies may provide greater insight into the physical processes
that shaped them (Bender, Burstein, \& Faber 1993).

The discovery of the fundamental plane relations was crucial
to the use of ellipticals in peculiar velocity surveys because of their
much increased accuracy over Faber-Jackson. For example,
Jorgensen, Franx, \& Kjaergaard (1996) estimate that
the scatter in $\log(R_e)$ at fixed $\sigma_e$ and $I_e$
is 0.084 dex. This corresponds to a distance error of
just over 19\%, quite comparable to recent estimates
of the TF distance error (\S\ref{sec:TF}). Moreover,
Jorgensen \etal\ have found that the distance error
is reduced to 17\% when galaxies with $\sigma_e<100\ \kms$
are excluded. This effect is reminiscent of the increased
TF scatter at lower velocity widths discussed in \S\ref{sec:TF} above,
and may arise for the same reason. Pahre, Djorgovski, \& de Carvalho (1995)
find a distance error of 16.5\% from the $K$-band FP. 

The \dnsigma\ relation occupies a special place in the history
of peculiar velocity surveys because it was used in the first
detection of very large-scale streaming by the ``7-Samurai''
group (Dressler \etal\ 1987;
Lynden-Bell \etal\ 1988).
In the 7-Samurai survey, a full-sky sample of elliptical galaxies 
revealed a streaming motion of amplitude \sm 500 \kms\ that was
coherent across the entire sky to a depth of \sm 40 Mpc. Subsequent
studies of spiral galaxies (Willick 1990; Han \& Mould 1992;
Mathewson \etal\ 1992; Courteau \etal\ 1993) have lent confirmation
to this result, although the coherence length of the flow remains
controversial. Since the late 1980s no new results concerning
the peculiar velocity field have been obtained using elliptical
galaxy data. However, this situation will change in the coming
years as several large surveys of elliptical galaxies (e.g., Wegner \etal\ 1996)
come to fruition.

Like the TF relation, the FP relations are now being studied
at appreciable redshift as well. Recently, Bender, Ziegler, \& Bruzual (1996)
have studied a sample of cluster ellipticals at $z=0.37.$ They
have found evidence for mild (\sm 0.5 mag) evolution
toward brighter magnitudes at such redshifts, comparable to 
the result found by Vogt \etal\ (1996) for the TF relation.

Because it is difficult to find Cepheids in nearby elliptical
galaxies, there has been little attempt to provide 
absolute calibrations of the \dnsigma\ and FP relations. As a result, 
elliptical galaxy distances have
not figured prominently in the Hubble constant problem.
However, this situation may change in the near future, if Surface Brightness Fluctuation
distances (discussed in the next Section)
can provide an absolute calibration for
the \dnsigma\ and FP relations.

\section{Surface Brightness Fluctuations}
\label{sec:SBF}
The last five years have seen the revival of an old
idea with modern technology: determining distances from
the ``graininess'' of a galaxy image. The basic idea
is simple. Galaxies are made up of stars. 
The discrete origin of galaxian luminosity
is detectable in the pixel-to-pixel 
intensity fluctuations of the galaxy image.
Such fluctuations derive from Poisson statistics of two sorts:
(1) photon number fluctuations $\left(\frac{\delta N}{N}\right)_\gamma,$
and (2) star number fluctuations $\left(\frac{\delta N}{N}\right)_s.$
The first 
is distance-independent, but $\left(\frac{\delta N}{N}\right)_s$
decreases with distance, as the solid angle
subtended by a pixel encompasses more and more individual stars.
Consequently, the pixel-to-pixel
intensity fluctuations in a nearby galaxy
are greater than in a more distant galaxy. If this
effect can be calibrated, it can be used as
a distance indicator.

Though originally proposed by Baum (1955), it was not
until the late 1980s that this idea has been put into practice,
made possible by the advent of CCD detectors 
and telescopes with improved
seeing. Tonry and coworkers (Tonry \& Schneider 1988;
Tonry, Ahjar, \& Luppino 1989,1990; Tonry \& Schecter 1990; Tonry 1991;
Tonry \etal\ 1997)
have pioneered this technique, which
has come to be known as the Surface Brightness Fluctuation
(SBF) method. The method can, in principle, be applied to
any type of galaxy. In practice, late-type ($\simgt$ Sb)
galaxies have too many sources of fluctuations over
and above Poisson statistics, such as spiral structure 
and dust lanes, to apply the method to them. The
method is thus preferentially applied to ellipticals and the
bulges of early-type spirals.

The distance at which SBF may be applied goes
inversely with the seeing. It is possible to measure distances
out to \sm 4000 \kms\ with a 2.4-meter telescope, \sm 2-hour
exposures, and
\sm 0.5 arcsecond seeing. This is an effective limit
for current ground-based observations. As half-arcsecond seeing
is infrequently achieved at even the best sites (such Mauna Kea 
and Las Campanas), 3000 \kms\ is a practical 
limit for complete SBF surveys. In principle, the HST is capable
of yielding SBF distances for objects as distant as 10,000 \kms.
However, the required exposure times are such that few galaxies
at such distances are likely to be observed for this purpose.

Tonry, Dressler, and coworkers have been conducting an SBF survey of \sm 400
early-type galaxies within \sm 3000 \kms\ over the last
six years (Dressler 1994; Tonry \etal\ 1997). 
As of this writing, the survey is nearly complete.
The data suggest that median SBF distance errors are \sm 8\% within
this distance range; the most well-observed objects have
distance errors of \sm 5\%. Such
accuracy is considerably better than most of the
secondary distance indicators discussed here, with the possible exception
of Type Ia supernovae (\S\ref{sec:SN}). 

\subsection{Calibration of SBF}
It may appear from the brief description above that
SBF is a purely ``geometrical method,'' like parallax.
If this were true,
it would free the method from the nagging questions
that plague other DIs: are they really universal, or
do they depend on galaxy type, age, environment, and
so forth? 
In reality, however, 
SBF is 
dependent on the stellar
populations in the galaxies to which it is applied.
Not only does this mean that we need to be cautious
with regard to its univerality, but, also,
that it is difficult to derive an absolute
calibration of SBF from first principles.
Like the other DIs considered in this Chapter,
absolute distances obtained from the SBF technique
are tied to the Cepheid distance scale. If the
latter were to change, so would the SBF distances.
In particular, estimates of $H_0$ derived from SBF
studies (see below) may well require revision
as the HST Key Project (\S\ref{sec:cepheid}) yields new results.

The stellar population dependence of SBF
arises because the stars which contribute most strongly
to the fluctuations are those that lie at the tip
of the giant branch. Tonry and coworkers parameterize
this effect in terms of ``effective fluctuation magnitudes''
$\overline{M_I}$ (absolute) and $\overline{m_I}$ (apparent). The quantity
$\overline{M_I}$ may be thought of as the absolute magnitude of 
the giant branch stars which dominate the fluctuations;
$\overline{m_I}$ is an apparent magnitude obtained from
the observed fluctuations. If all galaxies had
identical stellar populations, they would all
have the same value of $\overline{M_I},$ and their distance
moduli would be given simply by $\overline{m_I}-\overline{M_I}.$

\begin{figure*}
\centerline{\epsfxsize=3.2in \epsfbox{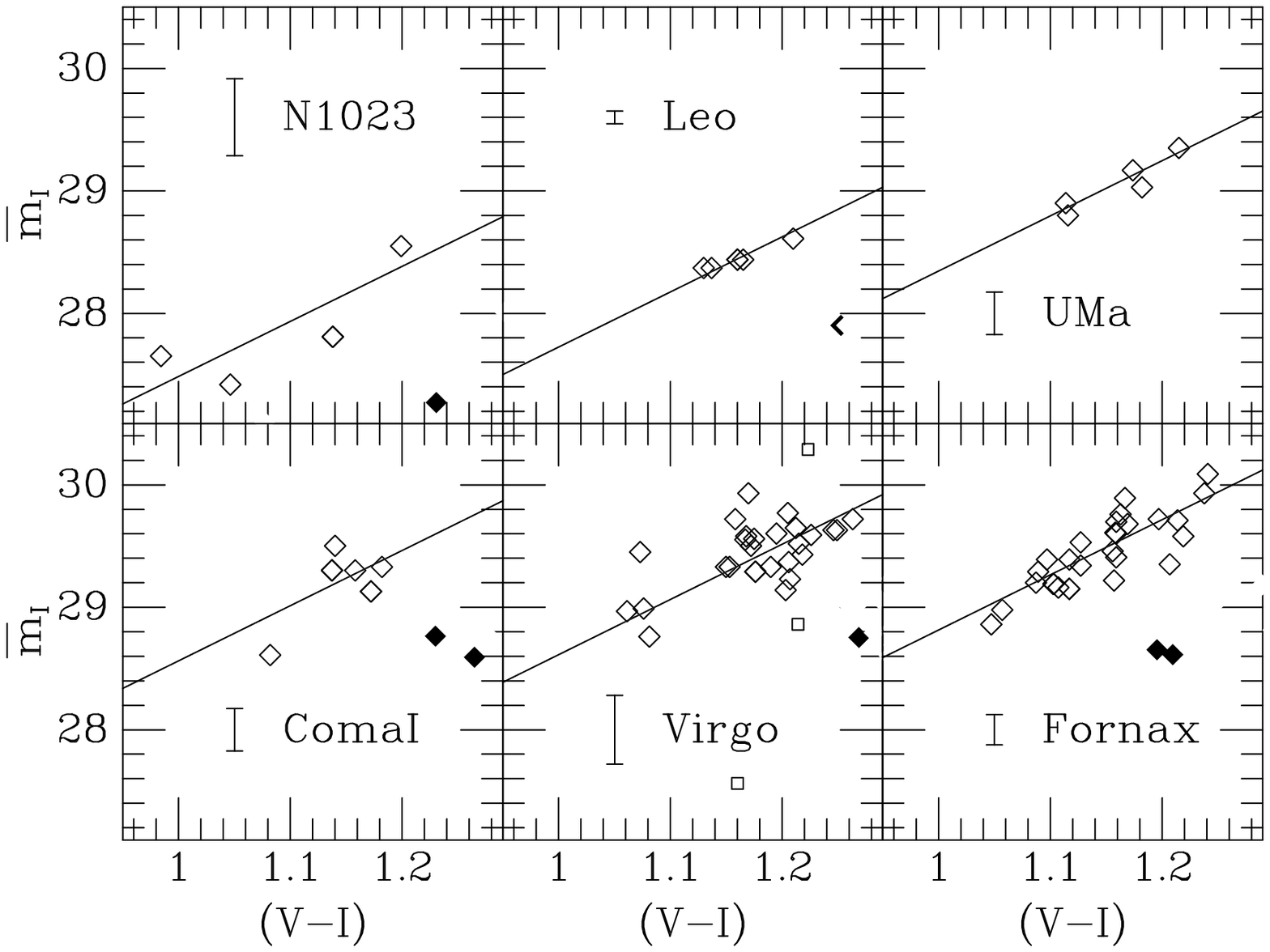}}
\centerline{\epsfxsize=3.0in \epsfbox{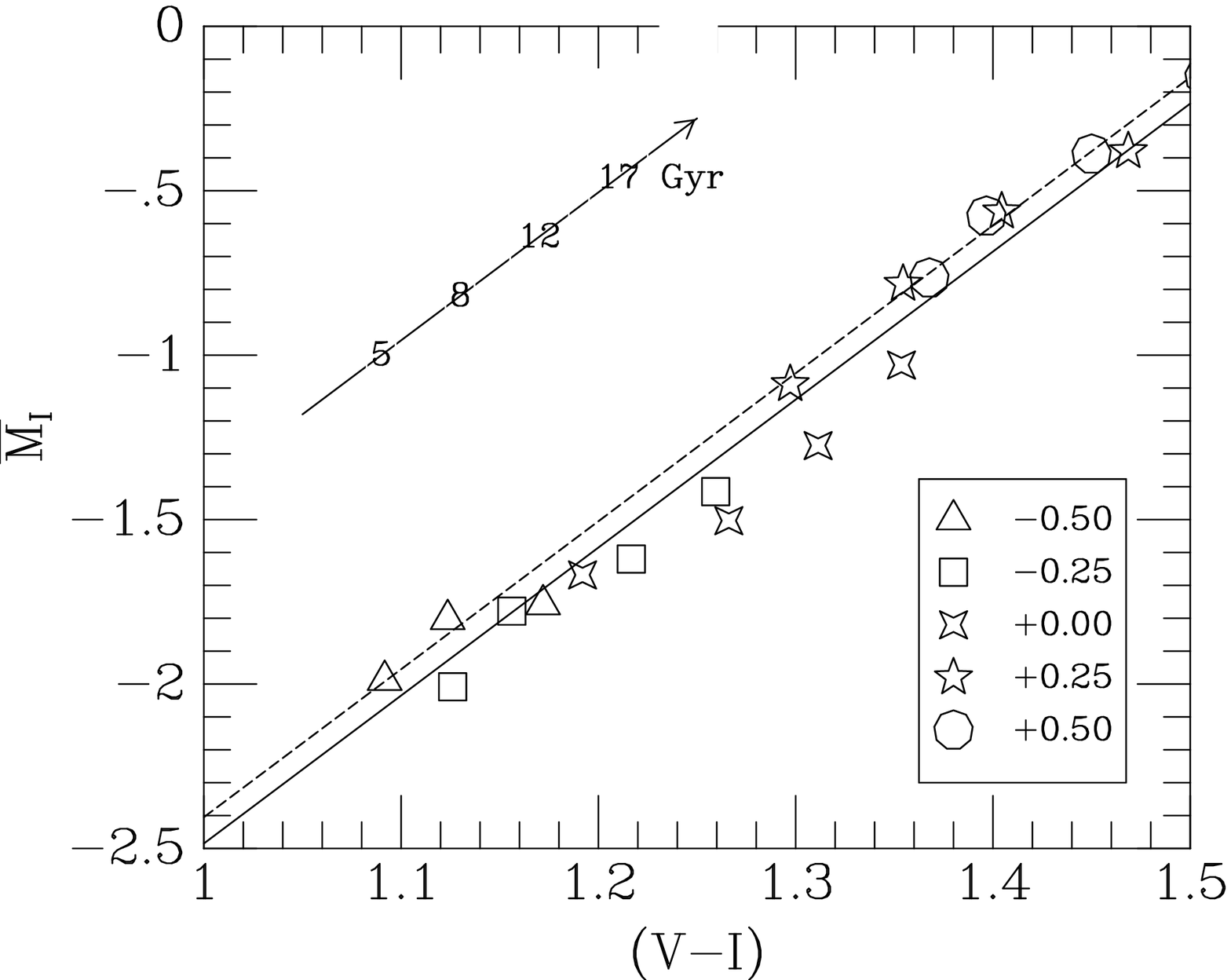}}
\label{fig:tonry_zpt}
\caption{Top panel: fluctuation apparent magnitudes $\overline{m_I}$
versus $(V-I)$ color for several nearby groups and clusters. The
solid lines drawn through the data points all have slope 4.5.
Bottom panel: a plot of the theoretical $\overline{M_I}$ versus
$(V-I)$ relation, from the stellar population synthesis models
of Worthey (1994). The different point types indicate different
metallicities relative to the Milky Way, as coded in the inset.
For each point type, there are several distinct points,
corresponding to different stellar population ages, as
indicated by the arrow. The solid line
is a fit to the theoretical models with
slope fixed at 4.5. The dashed line is the empirical relation,
equation~\ref{eq:sbf-calib}.
Adapted from Tonry \etal\ (1997).}
\end{figure*}

Because galaxies do not have identical stellar
populations, it is necessary to
determine an empirical correction to $\overline{M_I}$
as a function of a distance-independent galaxian
property. Tonry and coworkers use $(V-I)$ color
for this purpose. Their most
recent calibration is 
\begin{equation}
\overline{M_I}=-1.74(\pm 0.05) +4.5(\pm 0.25)\left[(V-I)-1.15\right]\,, 
\label{eq:sbf-calib}
\end{equation}
which is valid for $1.0\leq (V-I) \leq 1.3$ 
(Tonry \etal\ 1997). We discuss the zero point of this
relation in \S\ref{sec:sbf-results}. 
The color dependence indicated by equation~\ref{eq:sbf-calib}
is readily seen in the $\overline{m_I}$ versus $(V-I)$ diagrams 
of several tight groups and
clusters, as
shown in the upper panel of Figure~\ref{fig:tonry_zpt}.
A line of slope 4.5---the solid lines drawn
throught the data points---fits the fluctuation magnitude-color
data in each group well. (The solid points, as well
as the small open squares, are thought to be non-members of the groups.)
The different intercepts of the solid lines reflect the
different distances to the groups. 

The correlation of fluctuation magnitude
with color is quite strong. 
Accurate colors are therefore required in order to minimize
systematic effects. The
possibility that the slope or zero point of
this correlation may 
not be universal, but instead 
depend on some as yet undetermined galaxy properties
as suggested by Tammann (1992),
merits further attention. However, 
Tonry \etal\ (1997) show that the most
likely manifestation of such a problem, a trend
with metallicity of residuals from the $\overline{M_I}$--$(V-I)$
relation, does not exist.
It is reassuring, moreover,
that theoretical stellar population synthesis models
predict a trend of fluctuation magnitude with color
that is very similar to the empirical one.
This is shown in the lower panel of
Figure~\ref{fig:tonry_zpt}, in which the population synthesis models
of Worthey (1994) are plotted in the $\overline{M_I}$--$(V-I)$
plane. The points represent models of various metallicities
relative to the Milky Way (indicated by point type as
coded in the inset of the figure), and of various ages
(the trend with age, at a given metallicity, is indicated
by the arrow). The solid line, which has slope 4.5 and
an intercept determined by fitting to the
theoretical models, is seen to be a reasonable fit. The
dashed line is the emprical relation, equation~\ref{eq:sbf-calib}.
The zero point
of the theoretical relation differs from that of the empirical
one by only 0.07 mag. 

\subsection{Results from SBF Surveys}
\label{sec:sbf-results}
The ramifications of existing SBF data for
peculiar velocity surveys and the Hubble constant
are preliminary, but they are encouraging in terms of what they
portend for the knowledge this method will bring in the near future.
In the very early days
of peculiar velocity work, Tonry \& Davis (1980)
and Aaronson \etal\ (1982) estimated values of \sm 250 \kms\ for
the infall of the Local Group into the Virgo cluster.
Model fits to the SBF data for Local Supercluster
galaxies confirm this value, and show
that it is remarkably insensitive
to the assumed density profile around Virgo (Tonry 1995).
Another early scientific result of SBF studies
has been validation of the large peculiar motions
of elliptical galaxies in the Hydra-Centaurus region
originally detected using the \dnsigma\ technique
(Dressler 1994). More generally, intercomparison
of the SBF and TF/\dnsigma\ velocity fields in
the coming years will provide an important 
consistency check. Preliminary
tests of this sort have shown good agreement
to within the
quoted errors (Tonry 1995; Tonry \etal\ 1997). 

The zero point of the SBF method (i.e., the value of
$\overline{M_I}$ for a given $(V-I)$ color) was poorly known
until recently, but has now been determined
from a comparison of SBF and Cepheid distances.
Taking the distances in Mpc to the Local Group, the M81, CenA,
NGC 1023, NGC 3379, NGC 7331 groups, and the Virgo cluster
from published Cepheid data, Tonry \etal\ (1997) obtained
the zero point
given in equation~\ref{eq:sbf-calib}. By working with
groups, Tonry \etal\ were able to include 10 galaxies with
Cepheid distances and a total of 44 SBF galaxies 
in the calibration.
However, this comparison suffers from the nagging possibility
that the SBF objects, which are preferentially
ellipticals and S0s, may not lie at precisely the same
distances as the Cepheid galaxies, which are late-type
spirals, in the same group. Indeed, there are 
currently only five galaxies with
both Cepheid and SBF distances. One of these, NGC 5253,
gives a discordant result. If the remaining four are used,
Tonry \etal\ (1997) find an SBF zero point $-1.82\pm 0.06,$
in reasonable agreement with the preferred value of $-1.74\pm 0.05$
found from the group comparison.

Thus calibrated, the SBF technique can
be used as a temporary bridge between Cepheid distances, still
too few in number to be reliable calibrators, and the secondary
DIs that probe the far-field of
the Hubble flow. Tonry \etal\ (1997) used
SBF distances for groups and individual
galaxies to provide absolute calibrations for
TF, \dnsigma, and Type Ia supernovae (\S\ref{sec:SN}). In so doing,
they obtained distances in Mpc for relatively distant galaxies,
and thus estimates of the Hubble constant.
The mean value was found to be
$H_0=81\pm 6\ \kmsmpc$ from
SBF-calibrated secondary DIs. Such a large value of $H_0,$
if it holds up, may prove problematic for Big Bang cosmology,
as discussed in \S\ref{sec:cepheid}. However, it should be
kept in mind that the absolute calibration of SBF is tied
to the Cepheid distance scale, and that the latter might change
in coming years as the HST Key Project
(\S\ref{sec:cepheid}) continues.

\section{Supernovae}
\label{sec:SN}
The use of supernovae as distance
indicators has grown dramatically in the last few years. 
Supernovae have been applied
to the Hubble Constant problem, to measurement of
the cosmological parameters $\Omega_0$ and $\Lambda,$
and even, in a preliminary
way, to constraining bulk peculiar motions. There is every
reason to believe that in the next decade supernovae
will become still more important as distance indicators.
It is certain that many more will be discovered, 
especially at high redshift.

Supernovae come in two main varieties. Type Ia supernovae (SNe Ia) 
are thought to result from
the nuclear detonation of a white dwarf star that has been
overloaded by mass transferred from an evolved
(Population II) companion. (Recall that a white dwarf
cannot have a masss above the Chandrasekhar limit, $1.4 M_\odot.$
When mass transfer causes the white dwarf to surpass this
limit, it explodes.)
Type II supernovae result from the imploding cores of high-mass,
young (Population I) stars that have exhausted their nuclear 
fuel.\footnote{It is inconvenient that {\em Type I\/} supernovae occur in
{\em Type II\/} stellar populations, while {\em Type II\/}
supernovae occur in {\em Type I\/}
populations. Inconvenient nomenclature is, of course,
nothing new in astronomy---and must be tolerated as usual.} 
Of the two, it is the Type Ias that have received the most
attention lately. Type IIs have
shown somewhat less promise as distance indicators.
They are considerably fainter (\sm 2 mag), and thus
are detected less often in magnitude limited surveys
(although their intrinsic frequency of occurrence is in
fact greater than that of Type 1as).
The discussion
to follow will be restricted to Type Ias.

Because SNe Ias result (in all likelihood) from detonating white dwarfs, 
and because the latter tend to have very similar masses,
SNe Ias tend to have very similar luminosities.
That is, they are very nearly standard candles, so comparison
of their apparent and abolute magnitudes yields a distance.
Recent work suggests that Type Ia SNe are not quite
standard candles, in that their peak luminosities
correlate with the shape of their light curves (Phillips 1993;
Hamuy \etal\ 1995; Riess, Press, \& Kirshner 1995a,b;
Perlmutter \etal\ 1997). Basically, broad light curves
correspond to brighter, and narrow light curves to fainter,
supernovae. 
When this effect is accounted for, the
scatter in SNe Ia predicted peak magnitudes might be
as small as 0.1 mag, as found by Riess, Press, \& Kirshner (1995b).
Hamuy \etal\ (1995) and
Perlmutter \etal\ (1997)  
find that the scatter drops from
$\simlt 0.3$ mag mag when SNe Ia are treated as standard candles to 0.17
mag when the light curve shape is taken into account. 
The precise scatter of SNe Ias 
remains a subject for further study.

The wealth of new SNe data that has become available in
recent years is due to the advent of large-scale, systematic search techniques.
To understand this, it may be worth
stating the obvious. It is not possible to pick an arbitrary galaxy
and get a supernova distance for it because most
galaxies, at a given time, do not have a supernova in them.
Thus, it is necessary to search many galaxies at random
and somehow identify the small fraction ($\sim 10^{-4}$)
in which a supernova is going off at any given time.
Methods for doing this have been 
pioneered by Perlmutter and collaborators (Goobar \& Perlmutter 1995;
Perlmutter \etal\ 1995, 1996, 1997).  Deep images are taken of the same
region of the sky 2--3 weeks apart. Stellar objects which appear in
the second image but not in the first are candidate supernovae
to be confirmed by spectroscopy.
By means of such an approach, of order 30 high-redshift ($z=0.35$--0.65)
are now known. Related approaches for finding moderate- (Adams \etal\ 1995;
Hamuy \etal\ 1995) and high- (Schmidt \etal\ 1995) redshift 
supernovae have been developed by other groups as well. 

\begin{figure*}
\centerline{\epsfxsize=4.0 in \epsfbox{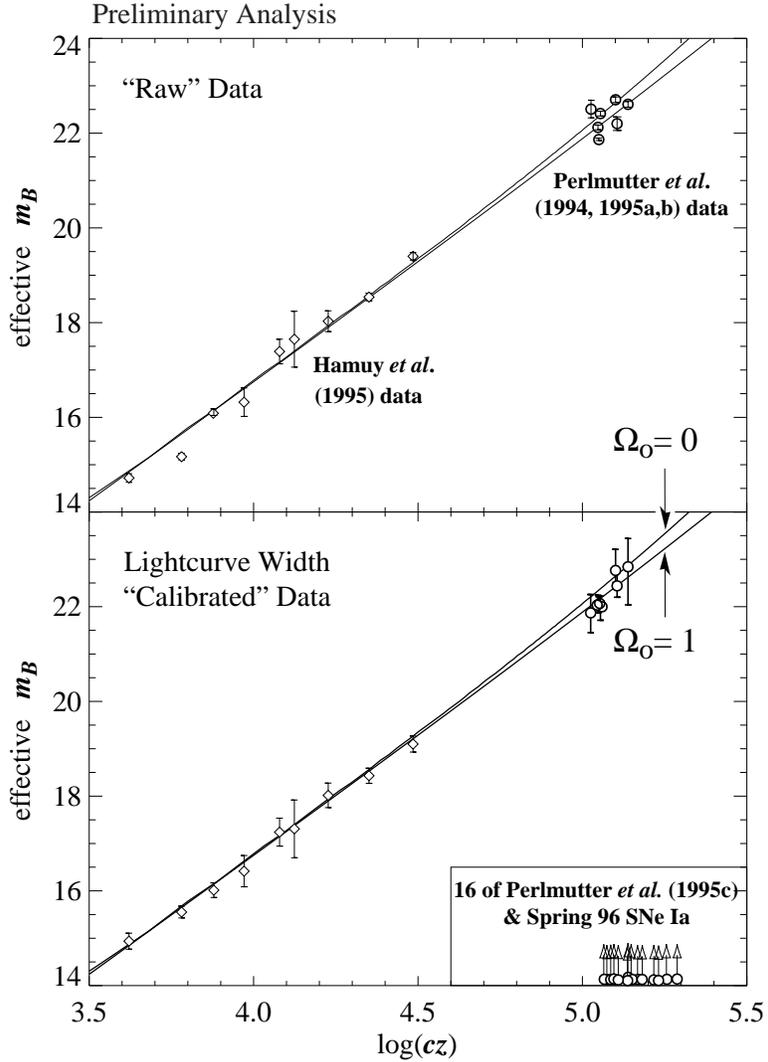}}
\caption{Hubble diagrams using SNe Ia. The
upper panel shows the observed peak apparent magnitudes;
in the lower panel the magnitudes are corrected for
the light curve width effect (see main text for details).
The inset in the lower panel shows the redshifts
of 16 additional SNe Ias
recently discovered but not yet analyzed by the Perlmutter group.}
\label{fig:perlmutter}
\end{figure*}
Search techniques such as those of the Perlmutter group
survey many faint galaxies in limited regions of the sky, and
are not very good at finding low-redshift ($z\simlt 0.03$) supernovae. Thus, they
are not particularly relevant to peculiar velocity studies
(but see below). However, precisely because they
detect intermediate to high redshift supernovae, such techniques
will be useful for measuring $H_0$ (with supernovae 
found at $z\simlt 0.2,$ where cosmological effects are
relatively unimportant),
and are among 
the best existing methods for
determining the cosmological parameters $\Omega_0$ and $\Lambda$
(with supernovae at $z\simgt 0.3,$ which probe spatial curvature.)
To see how this works, one can plot Hubble diagrams
for recently discovered supernovae both at moderate
and high redshift. This is done in Figure~\ref{fig:perlmutter},
which has been adapted from
the 1996 San Antonio AAS meeting contribution
by the Perlmutter group.
The low redshift data ($\log(cz)<4.5$) are from Hamuy \etal\ (1995),
and the high redshift data are from Perlmutter \etal\ (1996).

Figure~\ref{fig:perlmutter} contains several important features.
First, the observed peak apparent magnitudes are plotted versus log redshift
in the top panel.
To the degree SNe Ias are standard candles, one expects these
apparent magnitudes to go as ${\rm const.}+5\log(cz),$ 
the straight line plotted through the points at low redshift.
Correcting the SNe Ia magnitudes for the light
curve widths
(i.e., going from the top to the bottom panel) significantly
improves the agreement with this low-redshift prediction.
This is the main reason that the light curve width correction
is thought to greatly reduce the SNe Ia scatter.
Whether or not the correction is made, however,
the data provide unequivocal proof of
the linearity of the Hubble law at low ($z\simeq 0.1$) redshift.
Second, one expects that
that at higher redshifts the $m_B$-$\log(cz)$ relation will depart
from linearity because of spacetime curvature.
The departure from linearity is, to first order in $z,$ a function only
of the deceleration parameter $q_0$---or equivalently, if
the universe has vanishing cosmological constant $\Lambda$ (see below),
by the density parameter $\Omega_0,$ which in that case is exactly
twice $q_0.$ Figure~\ref{fig:perlmutter} assumes $\Lambda\equiv 0$
and thus labels the curves by $\Omega_0.$
There is a hint in the behavior of the light-curve-shape corrected
magnitudes that this departure
from linearity has been detected, and in particular that
$\Omega_0\simeq 1$ is a better fit to the data than
$\Omega_0 \simeq 0$
(Perlmutter \etal\ 1996). 

Neither $q_0$ nor $\Omega_0$ alone
fully characterizes
the departure from a linear Hubble diagram. 
More generally, the behavior of the Hubble diagram at high redshift
depends on the cosmological parameters
$\Omega_0$ and $\Omega_\Lambda\equiv \Lambda/3H_0^2.$
Perlmutter \etal\ (1997) suggest that the SNe Ia data should
be interpreted for now in the context of two cosmological
paradigms: a $\Lambda=0$ universe, and a spatially flat ($\Omega_0
+\Omega_\Lambda=1$) universe.\footnote{With a 
large sample of SNe Ias that spans a large redshift range,
it may be possible to constrain $\Omega_0$ and $\Omega_\Lambda$
separately, without assuming either a flat universe or
a vanishing cosmological constant (Goobar \& Perlmutter 1995).
The present data are not adequate for this purpose.}
Perlmutter \etal\ (1997) carried out a statistical
analysis of the
7 high-redshift ($0.354\leq z\leq 0.458$) supernovae
discovered in their survey, and the 9 lower redshift
SNe Ias found by the Hamuy group, that are shown
in Figure~\ref{fig:perlmutter}. They find that $\Omega_0
= 0.96^{+.56}_{-.50}$ if a $\Lambda=0$ universe is assumed.
If the universe is flat, $\Omega_0=0.98^{+.28}_{-.24},$ with
corresonding limits on $\Omega_\Lambda=1-\Omega_0.$
The constraints are stronger in the flat universe case because
of the strong effect of a cosmological constant on the apparent
magnitudes of high-redshift standard candles.
These results are, potentially,
highly significant for cosmology. Low-density, spatially flat models have become popular lately because
they make the universe older (for a given $H_0$ and $\Omega_0$),
provide a better fit to large-scale structure data than $\Omega_0=1$ models,
and yet remain consistent with the attractive idea that the early universe
underwent inflation. Currently favored versions of such models
have $\Omega_\Lambda \simeq 0.6$--$0.7$ (Ostriker
\& Steinhardt 1995). The SNe Ia results of
Perlmutter \etal\ (1997), which strongly disfavor such
a large $\Omega_\Lambda,$ will be difficult to
reconcile with low-density flat models. 

The analysis just described did not require absolute calibration
of SNe Ias. Indeed, Perlmutter \etal\ 1997 use a formalism
similar to that used in peculiar velocity studies, in which distances
are measured in \kms, and absolute magnitudes are, correspondingly,
defined only up to an arbitrary constant.
The SNe Ia data can be used to determine $H_0,$  however,
only to the degree
that the true absolute magnitudes (preferably corrected for
light curve width) of such objects are known. 
This requires either theoretical calibration
or empirical calibration in galaxies with
Cepheid distances. Both of these approaches pose difficulties. 
A range of models of exploding white dwarfs predict peak
absolute magnitudes for SNe Ias of
$M_V\simeq -19.5$ with small scatter, but significantly lower luminosities
can result if some of the key inputs to the models (especially the
mass of the $^{56}$Ni ejectae) are varied (H\"oflich, Khokhlov, \& Wheeler 1995).
This suggests that 
the absolute magnitudes of SNe Ias cannot yet be predicted
theoretically, and that an
empirical calibration using Cepheid distances
will do better. However, because local galaxies with Cepheid distances
are scarce, and SNe Ias are rare, there are still few reliable local
calibrators for SNe Ias. It has been necessary to analyze historical
as well as modern
SNe Ia data (Saha \etal\ 1995; Sandage \etal\ 1996) 
in Cepheid galaxies in order to
increase the number of calibrators. This approach encounters
the problem of relating modern CCD photometry with photometric methods
from decades past. Pending the detection and analysis of SNe Ias in 
a larger number of local galaxies with Cepheid distances, one should
view estimates of $H_0$ inferred from supernovae as preliminary.

Being rare events,
SNe Ias are unlikely to provide a detailed map of the local peculiar velocity field.
However, because of their small scatter (see above), a few
well-observed SNe Ias distributed on the sky  may lead to useful constraints on
amplitude and scale of large-scale bulk flows. 
A first attempt at this was carried out
by Riess, Press, \& Kirshner (1995b), who used 13 SNe Ias with peak
magnitudes corrected by light curve widths to place limits on the
bulk flow within \sm 7000 \kms. They found the data to be consistent
with at most a small ($\simlt 400\ \kms$) bulk streaming, and to
be inconsistent with the large bulk flow
found by Lauer \& Postman (1994) using an independent method (cf.\
\S\ref{sec:BCG} below). However, one must be cautious in interpreting
such results because small-scale power in the velocity field can
obscure large-scale motions 
(Watkins \& Feldman 1995).
Constraints on bulk
flows using SNe Ias are likely to improve in the coming years. 

\section{Brightest Cluster Galaxies}
\label{sec:BCG}
Another ``classical'' distance indicator method that has been reborn in
modern guise is photometry of brightest cluster galaxies (BCGs). As originally
treated by Sandage and coworkers (Sandage 1972; Sandage
\& Hardy 1973), BCGs were considered
to be good standard candles. As such, they were used to demonstrate
the linearity of the Hubble diagram to relatively large distances 
and estimate $H_0.$ Any such estimate was and remains highly
suspect, however, because of the difficulty of obtaining a good absolute
calibration of the method. The scatter of BCGs as standard
candles is around 0.30--0.35 mag, which compares
favorably with methods such as TF or \dnsigma.

\begin{figure*}
\centerline{\epsfxsize=4.0 in \epsfbox{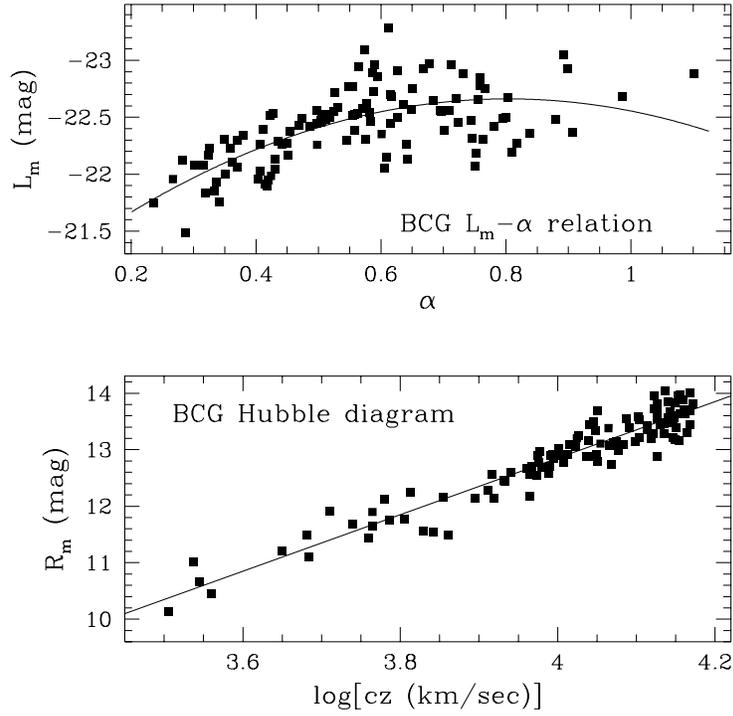}}
\caption{Top panel: the BCG $L_m-\alpha$ relation exhibited by the
sample of Lauer \& Postman (1992). Absolute magnitude within
the metric radius $r_m$ is plotted against the logarithmic
surface brightness slope at $r_m.$ 
The solid curve shows
the quadratic fit to the data. 
Bottom panel: the Hubble diagram for the Lauer \& Postman (1992)
BCG sample. Apparent magnitude within $r_m$ is plotted against
log redshift. The straight line plotted through the points has slope 5,
the relation expected for a linear Hubble flow.
The data used
to make this figure were kindly provided by Marc Postman.}
\label{fig:BCG}
\end{figure*}

A dubious assumption in the early work was that BCGs
are true standard candles. Gunn \& Oke (1975) first
suggested that the luminosities of BCGs 
might correlate with their surface brightness profiles. Following
this suggestion, Hoessel (1980)
defined
a metric radius $r_m=10 h^{-1}$ kpc, and showed that 
the metric luminosity $L(r_m) \equiv L_m$ varied roughly linearly
with a shape parameter $\alpha$ defined by
\begin{equation}
\alpha\equiv \left.\frac{d\log L}{d\log r}\right|_{r_m}\,.
\label{eq:defalpha}
\end{equation}
More recently, Lauer \& Postman (1992) have shown that the correlation
between $L_m$ and $\alpha$ is better modeled by a quadratic relation. 
The Lauer \& Postman (1992) data, along with their quadratic
fit, are shown in the upper panel of Figure~\ref{fig:BCG}. Thus
modeled, the typical distance error incurred by the BCG $L_m$-$\alpha$
relation is \sm 16\%.

A slight hitch in applying the BCG $L_m$-$\alpha$ relation is the requirement
of defining a metric radius $r_m$ for evaluating both $L_m$ and $\alpha.$
This means that the assumed peculiar velocity of a BCG must be factored
in to convert redshift to distance, and thus angular to linear diameter.
In practice this is not a very serious issue. At the typically large distances
($\simgt 7000\ \kms$) at which the relation is applied, peculiar velocity
corrections have a small effect on $L_m$ and $\alpha.$ Iterative techniques
in which a peculiar velocity solution is
obtained and then used to modify the $r_m$s, converge quickly
(Lauer \& Postman 1994). 

Modern scientific results based on BCGs are due to
the pioneering work of Lauer and
Postman (Lauer \& Postman 1992; Lauer \& Postman 1994, hereafter LP94;
Postman \& Lauer 1995).
One important---and uncontroversial---such result 
has been confirmation, with unprecedented accuracy, of the linearity of the Hubble
diagram to redshifts $z\simeq 0.05$ over the entire sky. (The
Hubble diagrams using SNe Ias (\S\ref{sec:SN}), by contrast,
are not derived from isotropic samples.) 
This is
shown in the lower panel of Figure~\ref{fig:BCG}.
However, another
result has been considerably more controversial,
namely, the detection of a very large-scale bulk peculiar velocity by LP94.
The linearity of the BCG Hubble diagram manifests
itself with the smallest scatter when the velocities are
referred to a local frame that differs significantly from that defined
by the CMB dipole. Or, stated another way, the LP94 data indicate that the
frame of Abell clusters out to 15,000 \kms\ redshift is moving with
respect to the CMB frame at a velocity of \sm 700 \kms\ toward
$l\simeq 350\degs,$ $b\simeq 50\degs.$ A reanalysis of the LP94 data by
Colless (1995) produced a very similar result for the bulk motion.

The global Hubble flow linearity 
demonstrated by Lauer \& Postman (1992) suggests that that the BCG
$L_m$-$\alpha$ relation is an excellent DI
out to substantial redshifts. However, the indicated bulk
motion is of sufficient amplitude and scale
as to appear inconsistent with other indicators of
large-scale homogeneity. For example, Strauss \etal\ (1995) showed
that none of the leading models of structure formation
that are consistent with other measures of
large-scale power can reproduce an LP94-like result in more than a small
fraction of realizations. Furthermore, two recent studies, one using the
TF relation (Giovanelli \etal\ 1996) and one using Type Ia SNe
(Riess, Press, \& Kirshner 1995b), suggest that the bulk motion on 
smaller scales than that probed by the BCGs is inconsistent with the
LP94 bulk flow at high significance levels.

For the above reasons, the current status of BCGs as DIs is controversial.
However, one should not prejudge the outcome. Velocity studies have yielded
a number of surprises in the last 15 years, and it is not inconceivable
that the LP94 bulk flow---or something like it---will be vindicated in
the long term. Lauer, Postman, and Strauss are currently extending BCG
observations to a complete sample with $z\leq 0.1,$ and the results of
their survey are expected to be available by \sm late 1997. Whether 
or not it 
confirms LP94, this extended study is likely to greatly
clarify the nature of the BCG $L_m$-$\alpha$ relation.

\section{Redshift-Distance Catalogs}
\label{sec:catalogs}

As redshift measurements accumulated in the 1970s and 1980s,
it was widely recognized that there was a need to assemble these
data into comprehensive catalogs. Beginning with the publication
of the CfA redshift survey in 1983 (Huchra \etal\ 1983),
all major redshift surveys (see the Chapter by Strauss
in this volume) led to electronically available
databases in fairly short order. 

Comparable efforts involving
redshift-independent distance measurements have been slower
in coming. This is largely due to the issue of {\em uniformity.}
Whereas redshift measurements by different observers rarely
exhibit major differences, redshift-independent
distances obtained by different observers can, and generally do,
differ systematically for any number of reasons. In some cases
the origin of such differences is different calibrations of the
DI. In others, the calibrations are the same but the input data
differ in a subtle way. Finally, the way statistical bias effects
are treated (\S\ref{sec:bias}) often differs among those
involved in galaxy distance measurements. For all these reasons,
it is not possible simply to go to the published literature, find
all papers in which galaxy distances are reported, and
lump them together in a single database. Instead, individual
data sets must be assembled, their input data and selection
criteria characterized, their DI relations recalibrated if necessary,
and the final distances brought to a uniform system. Only
then can the resultant catalog be relied upon---and even then,
caution is required.

The first steps toward assembling homogeneous redshift-distance
catalogs were taken in the late 1980s by David Burstein.
His goal was to combine the then newly-acquired \dnsigma\ data
from the 7-Samurai group (\S\ref{sec:dnsigma}) with the extant
data on spiral galaxy distances, especially the infrared
TF data obtained by the Aaronson group (\S\ref{sec:TF}). 
Burstein's efforts produced two electronic
catalogs, the Mark I (1987) and Mark II (1989) Catalogs of Galaxy Peculiar
Velocities.\footnote{Although these are referred to as ``peculiar velocity''
catalogs, they are, first and foremost, {\em redshift-distance\/} catalogs,
consisting of redshifts and redshift-independent distances. The peculiar
velocities follow from these more basic data, although not necessarily
in a simple way, given the statistical bias effects 
studied in \S\ref{sec:bias}.}
Burstein's chief
concern was matching the TF and \dnsigma\ distance scales. As there 
are, by definition, no galaxies that have both kinds of distances,
this matching could be carried out through a variety of overlapping
approaches. The approach decided upon by Burstein, in consultation
with the other 7 Samurai, was to require the Coma cluster spirals
and ellipticals to have the same mean distances. 
Although this procedure was imperfect, the Mark II catalog
was considered reliable enough to be
used in the first major effort to constrain the density parameter
$\Omega_0$ by comparing velocities and densities (Dekel \etal\ 1993).

With the publication of a number of large, new TF data sets in the early
1990s, the need for a greatly expanded redshift-distance catalog became
apparent. An important development was the superseding of the majority of the
older infrared TF data, obtained by the Aaronson group, by CCD-based
($R$-and $I$-band) TF data. Han, Mould and coworkers (Han 1991, 1992;
Han \& Mould 1992; Mould \etal\ 1991, 1993) obtained a full-sky cluster
TF sample, based on $I$ -band magnitudes and 21 cm velocity widths,
comprising over 400 galaxies. Willick (1990, 1991) and Courteau (1992;
Courteau \etal\ 1993) gathered $R$-band TF data in the Northern
sky for over 800 galaxies in total. The largest single contribution was
that of Mathewson, Ford, \& Buchorn (1992) who published an $I$-band
TF sample of 1355 galaxies in the Southern sky. Despite
the influx of the new CCD data, one portion of the infrared TF database of
the Aaronson group was not rendered obsolete: the sample of over
300 local ($cz\simlt 3000\ \kms$) galaxies first observed in
the late 1970s and early 1980s (Aaronson \etal\ 1982). This
local sample was, however, subjected to a careful reanalysis by
Tormen \& Burstein (1995), who 
rederived the $H$-band magnitudes
using a more homogeneous set of galaxy diameters and
inclinations than was available to the original researchers a decade earlier.

In 1993, a group of astronomers (myself, Burstein, Avishai Dekel,
Sandra Faber, and St\'ephane Courteau) began the process
of integrating 
these TF data and the existing \dnsigma\ data into a new
redshift-distance catalog. Our methodology is described in detail
in Willick \etal\ (1995, 1996), and portions of the catalog are  presented
in Willick \etal\ (1997). The full catalog, known as the
{\em Mark III Catalog of Galaxy Peculiar Velocities,}
is quite large (nearly 3000 spirals and over 500 ellipticals,
although this includes several hundred overlaps between data sets)
and is available only electronically, as described in Willick \etal\ (1997).

Building upon the foundation laid by Burstein in the Mark I and II catalogs,
the Mark III catalog was assembled with special emphasis placed on achieving
uniform distances among the separate samples it comprises. Four specific
steps were taken toward this goal. First,
the raw data in all of the TF samples underwent
a uniform set of corrections for inclination and extinction (cf.\
\S\ref{sec:TFdetails}). Second, the TF relations for each sample
were recalibrated using a self-consistent procedure that included
correction for selection bias (\S\ref{sec:bias}). Third,
final TF zero points were assigned by requiring that the TF distances
of objects common to two or more samples agree in the mean. This
step ensures that the different samples are on similar relative
distance scales. The global TF zero point was determined by the
fully-sky Han-Mould cluster TF sample. (As explained in \S\ref{sec:TF},
this zero point was such that the distances are given in units of \kms, not Mpc.)
Fourth, the spiral and elliptical distance scales were matched by
applying the POTENT algorithm (see the chapter by Dekel in this volume)
to each separately, and requiring that they produce statistically 
consistent velocity fields.

In parallel with the efforts of the Mark III group, similar enterprises
have been undertaken by two other groups. Brent Tully has also
assembled and recalibrated much of the extant TF data. Riccardo Giovanelli,
Martha Haynes, Wolfram Freudling, Luiz da Costa, and coworkers have 
acquired new $I$ band TF data for
\sm 2000 galaxies, and have combined it with the sample of
Mathewson \etal\ 1992). Initial scientific results from each of
these efforts have been published (Shaya, Tully, \& Peebles 1995;
Giovanelli \etal\ 1996; da Costa \etal\ 1996), and the catalogs
themselves will soon become publically available.

New distances for elliptical galaxies, now mostly from
the FP rather than \dnsigma\ (\S~\ref{sec:dnsigma}),
continue to be obtained as well. Jorgensen, Franx, \& Kjaergaard (1995a,b)
have published distances for E and S0 galaxies in 10 clusters out
to 10,000 \kms. The EFAR group (Burstein, Colless, Davies, Wegner,
and colleagues) are now finishing an FP survey of over 80 groups
and clusters at distances between 7000 and 16,000 \kms\ (Colless \etal\ 1993;
Wegner \etal\ 1993,1996; Davies \etal\ 1993).

Implicit in all this ongoing work is that the Mark III catalog,
like its predecessors, is just one step along a path still
being traveled. Just as the Mark III data consists in part
of recalibrated data alreay present in the Mark II, so
will future catalogs incorporate, partially recalibrate,
and expand upon the Mark III.
Of particular note are the distances coming from
the SBF survey of Tonry and coworkers
(Tonry \etal\ 1997; cf.\ \S\ref{sec:SBF}). 
The SBF distances are much more accurate
than either TF or \dnsigma\ and can provide important checks
on them. Tonry \etal\ (1997) have taken initial steps toward 
such an intercomparison, and the preliminary results, which
suggest mutally consistent results among SBF, \dnsigma, and
TF, are encouraging. Little comparison of SNe and BCG distances
with other DIs has yet been carried out, but will be in the coming
years. 
It is reasonable
to hope that, by the turn of the century at the latest, the available
redshift-distance catalogs will be superior, in terms of
sky coverage, accuracy, and homogeneity, to the best we have today.

\section{Malmquist and Other Biases}
\label{sec:bias}
Distance scale and peculiar velocity work have long been plagued
by statistical biases. These biases are sufficiently confusing
and multifaceted that their effects are often
misunderstood or misrepresented. It is worth taking a moment
to go over a few of the main issues.

The root problem is that our distance indicators contain
scatter: a galaxy with distance $d$ inferred from the DI
really lies within some range of distances, approximately
(but not exactly) centered on $d.$ This range is characterized
by a non-gaussian
distribution of characteristic width $d\Delta,$ where
$\Delta$ is the fractional distance error characteristic
of the DI. (If $\sigma$ is the DI scatter in magnitudes,
$\Delta \simeq 0.46 \sigma$). Thus,
the farther away the object is the bigger the distance error.
For most DIs, a good approximation is that
the distribution of distance errors is
log-normal: if the true distance is $r,$
then the distance estimate $d$ has a probability distribution
given by 
\begin{equation}
P(d|r)=\frac{1}{\sqrt{2\pi}(d\Delta)}\exp\left[-\frac{\left[\ln(d/r)\right]^2}
{2\Delta^2}\right]\,.
\label{eq:pdgr}
\end{equation}

Two distinct kinds of statistical bias effects can arise when
DIs with the above properties are used.
Which of the two occurs depends on which of two
basic analytic approaches one adopts for treating
the DI data. In the first approach, known as {\em Method I,}
one assumes that the DI-inferred distance $d$ is the
best \apriori\ estimate of true distance. Any subsequent 
averaging or modeling of the data points assumes galaxies
with similar values of $d$ to be neighbors in real space
as well. The second approach, known as {\em Method II,}
takes proximity in redshift space as tantamount to
real-space proximity; the DI-inferred distances
are then treated only in a statistical sense, averaged
over objects with similar redshift-space positions.
The Method I/Method II terminology
originated with Faber \& Burstein (1988); a detailed discussion
is provided by Strauss \& Willick (1995, \S 6.4).

Let us consider this distinction in relation
to peculiar velocity or Hubble constant studies.
In a Method I approach, one would take
objects whose DI-inferred
distances are within a narrow range of some value $d,$ 
and average their redshifts. Subtracting
$d$ from the resulting mean redshift yields a peculiar velocity estimate;
dividing the mean redshift by $d$ gives an estimate of $H_0.$
However, these estimates will be biased, because the distance
estimate $d$ itself is biased:
{\em It is not the mean true distance of the objects in question.}
To see this, we reason as follows: if $P(d|r)$ is given
by equation~(\ref{eq:pdgr}) above, then the distribution of
true distances of our objects is given, according to Bayes' Theorem, by
\begin{equation}
P(r|d)=\frac{P(d|r)P(r)}{\int_0^\infty P(d|r)P(r)\,dr}
= \frac{r^2 n(r) \explnrD}{\int_0^\infty r^2 n(r) \explnrD\,dr}\,,
\label{eq:prd}
\end{equation}
where we have taken $P(r)\propto r^2 n(r),$ where $n(r)$ is the
underlying galaxy number density along the line of sight.
To obtain the expectation value of the
true distance $r$ for a given $d,$ we
multiply equation~(\ref{eq:prd})
by $r$ and integrate over all $r.$ In general, this
integral requires knowledge of the density field $n(r)$
and will have to be done numerically. However, in the
simplest case that the density field is constant, the
integral can be done analytically. The result is
that the expected true distance is $de^{7\Delta^2/2}$
(Lynden-Bell \etal\ 1988; Willick 1991).
This effect is called {\em homogeneous Malmquist bias.} It tells
us that, typically, objects lie further away than their
DI-inferred distances. The physical cause
is more objects ``scatter in''
from larger true distances (where there is more volume)
than ``scatter out'' from smaller ones. In general, however, variations
in the number density cannot be neglected. When this is
the case, there is {\em inhomogeneous Malmquist bias} (IHM).
IHM can be computed numerically if one has a model of 
the density field. Further discussion of this issue
may be found in Willick \etal\ (1997).

The biases which arise in a Method II analysis are quite different.
They may be  rigorously understood in terms of the probability distribution
of the DI-inferred distance $d$ given the redshift $cz,$
$P(d|cz)$ (contrast with equation~\ref{eq:prd}, which
underlies Method I). In general, this distribution
is quite complicated (cf.\ Strauss \& Willick 1995, \S 8.1.2),
and its details are beyond the scope of this Chapter.
However, under the assumption of a ``cold'' velocity field---an
assumption that appears adequate in ordinary environments---redshifts
complemented by a flow model
give a good approximation of true distance. Thus, it really is the
probability distribution $P(d|r)$ (equation~(\ref{eq:pdgr}), or
one similar to it,
that counts for a Method II analysis. However, that equation as written
does not represent the full story. If severe
selection effects such as a magnitude or diameter limit
are present, then the log-normal distribution does not apply exactly.
Some galaxies are too faint or small to be in the
sample; in effect, the large-distance tail of $P(d|r)$
is cut off. It follows that 
the typical inferred distances are {\em smaller\/} than
those expected at a given true distance $r.$ 
As a result, the peculiar velocity model that
allows true distance to be estimated as a function of redshift is
tricked into returning shorter distances. This bias goes
in the same sense as Malmquist bias, but is fundamentally
different. It results not from volume/density effects,
but from {\em sample selection\/} effects, and
is called {\em selection bias.} 

Selection bias can be avoided, or at least minimized, by
working in the so-called ``inverse direction.'' What that
means is most easily illustrated using the
TF relation. When viewed in its ``forward'' sense,
the TF relation is conceived as a prediction of
absolute magnitude given a value of the velocity
width parameter, $M(\eta).$ However, it is equally
valid to view the relation as a prediction of $\eta$
given a value of $M,$ i.e., as a function $\eta^0(M)$
(the superscript ensures that there is no confusion
between the observed width parameter $\eta$ and the
TF-prediction). When one uses the forward relation,
one imagines fitting a line $m_i=M(\eta_i)+\mu$ by regressing
the apparent magnitudes $m_i$ on the velocity widths $\eta_i$; 
the distance modulus $\mu$
is the free parameter solved for. Selection bias then
occurs because apparent magnitudes fainter than
the magnitude limit are ``missing'' from the
sample, so the fitted line is not the same
as the true line. However, if one instead fits a line
$\eta^0(m_i-\mu)$ by regressing the widths on the
magnitudes, the same effect does not occur, provided
the sample selection procedure does not exclude large
or small velocity widths. In general, this last caveat
is more or less valid. Consequently, working in the
inverse direction does in fact avoid or at least minimize
selection bias.

This fact, first clearly stated by Schechter (1980) and
then reiterated in various forms by Aaronson \etal\ (1982),
Tully (1988), Willick (1994), Dekel (1994), and Davis, Nusser, \& Willick
(1996), among others, remains an obscure one, not universally
appreciated. It is often heard, for example, that the TF
relation applied to relatively distant galaxies will necessarily
result in a Hubble constant that is biased high, because
the distances are biased low due to selection bias.
The clear conclusion
of the previous paragraph, however, is that provided the
analysis is done using redshift-space information to assign \apriori\
distances---that is, provided that a {\em Method II\/}
approach is taken---working in the inverse direction can render selection
bias unimportant. It is also the case that a careful analytical
methods (Willick 1994) can permit a correction for selection
bias even when working in the forward direction. It should
be borne in mind, however, that both of these approaches
(using the inverse relation or correction for forward selection
bias) necessitate a careful characterization of sample
selection criteria. 

\begin{table}[t]
\centerline{\begin{tabular}{l||l|l}
\multicolumn{3}{c}{``Method Matrix'' of Distance Indicator Biases} \\ \hline\hline
{\large DI type/Method Type} & {\bf{\large Method I}} & {\bf{\large Method II}} \\
& {\em DI-inferred distance best} & {\em Redshift-space data best} \\
& {\em indicator of true distance} & {\em indicator of true distance} \\ \hline\hline
{\bf{\large Forward}} & {\bf Malmquist bias\/} & {\bf Strong selection bias\/} \\
{\em dist-dep (e.g.\ mag) predicted} & (selection-independent & (depends on observation
al\\
{\em by dist-indep (e.g.\ $\eta$) quantity} &  &
selection criteria) \\ \hline
{\bf{\large Inverse}} & {\bf Malmquist bias\/} & {\bf Weak or no selection bias\/} \\
{\em dist-indep predicted} & (selection-dependent) & (bias present if selection \\
{\em by dist-dep quantity} &  & related to dist-indep quantity) \\ \hline\hline
\end{tabular}}
\medskip
\label{matrixtable}
\end{table}

Another wrinkle in this complicated subject is that the
relatively bias-free character of inverse distance indicators
does not carry over to a Method I analysis. It is beyond the
scope of this Chapter to discuss this issue in full detail;
the interested reader is referred to Strauss \& Willick (1995,
\S~6.5). The main point is that a Method I inverse DI analysis
is subject to Malmquist bias in much the same way as a
Method I forward analysis; indeed, the inverse Malmquist
bias is in some ways considerably more complex, as it
depends (unlike forward Malmquist bias)
on sample selection criteria. So while it is correct
to emphasize the bias-free (or nearly so) nature of
working in the inverse direction, it is essential
to remember that this property holds only for Method
II analyses. 

Much of the confusion surrounding the
relative bias properties of forward versus inverse
DIs stems from neglecting the distinction between
Method I and Method II analyses.
Recognizing this, Strauss \& Willick
(1995) summarized the issue with what they called
the ``Method Matrix'' 
(a more memorable term might be the ``magic square'')
of peculiar velocity analysis.
Their table is reproduced above, in a slightly simpler
form (the original alluded to several complications
that are unecessary here).
Reference to this simple diagram might allay some of the
controversies surrounding Malmquist and related biases.

\section{Summary}
\label{sec:summary}
The measurement of galaxy distances is crucial for some of the
basic problems in astronomy and cosmology. In this Chapter I
have emphasized the role such measurements play in two of the most important:
Hubble constant determination and peculiar velocity 
analysis. An important distinction between these two efforts, which
I have reiterated throughout, is that for peculiar velocities one only
needs distances in \kms, which are independent of
an absolute distance scale, whereas for determination of $H_0$
distances in Mpc are required. In practice, this means that peculiar
velocity studies may be carried out using distance indicators such
as TF or \dnsigma\ calibrated only relative to the distant Hubble
flow. To obtain $H_0$ the same DIs must be calibrated relative to
local galaxies with Cepheid distances. Because the program of Cepheid
measurements in local calibrators using HST (Kennicutt \etal\ 1995)
is ongoing, reliable far-field measurements of $H_0$ are still several
years away. 

I have organized the discussion around the principal distance indicators
currently in use. These are:
\begin{enumerate}
\item Cepheid variables. The Period-Luminosity relation for
these pulsating stars may be calibrated in the Milky
Way and in the Magellanic Clouds. However, they are detectable with HST out
to \sm 20 Mpc. As such, they will yield accurate absolute distances
for \sm 20 local galaxies over the next several years. These local
galaxies will in turn provide absolute calibrations for the secondary
distances indicators such as TF or SNe Ia that will be used
to measure $H_0$ in the ``far field'' ($\simgt 7000\ \kms$),
where peculiar velocities and depth effects are relatively unimportant.
\item The TF relation. This method has been the workhorse of peculiar
velocity studies, for it applies to the ordinary spiral galaxies that best
trace the peculiar velocity field. When calibrated using HST Cepheid
distances, it promises also to yield a value of $H_0$ accurate to
\sm 10\%. The TF relation has recently been shown to apply to
spiral galaxies at high redshift (Vogt \etal\ 1996), although evolutionary
effects appear to be significant at $z\simeq 0.5.$
\item The \dnsigma\ relation. This is a variant of the Fundamental Plane
relations for elliptical galaxies. It is comparable to TF in accuracy, and
gives similar global results for the large-scale peculiar velocity field
(Kolatt \& Dekel 1994). Its best chance for absolute calibration comes
from a comparison with SBF distances. Like TF, \dnsigma\ has recently
been applied to relatively high redshift galaxies (Bender \etal\ 1996),
again with evidence of evolutionary changes.
\item Surface Brightness Fluctuations (SBF). This method may be the most
accurate DI known for galaxies beyond the range of HST Cepheid
measurements, with distance errors as small as 5\% under the best
conditions and median errors of \sm 8\%. Its application is most straightforward
for early-type systems, although with care it may be applied to spirals
as well. It holds the promise of giving a high-resolution picture of the
peculiar velocity field. It will also provide a crucial check of the reliability
of TF and \dnsigma. Its direct application to the $H_0$ problem remains uncertain
because of the great technical challenge involved in extending it to
distances $\simgt 5000\ \kms.$
\item Type Ia Supernovae. SNe are in principle excellent DIs, but suffer from
the obvious problem that one cannot, in general, be found in
a given galaxy at a given time. In recent years, improved search techniques
have vastly increased the number of well-observed SNe Ias, both at relatively
low (Hamuy \etal\ 1995) and high (Perlmutter \etal\ 1996) redshifts. The
results of these studies have included beautiful Hubble diagrams that
demonstrate the linearity of the Hubble expansion to $z\simeq 0.1,$ with
tantalizing hints of curvature that hold the promise of constraining the
cosmological parameters $\Omega_0$ and $\Lambda.$
Sandage and coworkers (Sandage \etal\ 1996;
Saha \etal\ 1995) have calibrated SNe Ias in galaxies with Cepheid distances
to obtain Hubble constant estimates of $H_0\simeq 57\ \kmsmpc.$ However,
considerable uncertainty attaches to these results at present. The quest for
a reliable absolute calibration of SNe Ias continues.
\item The BCG $L_m$-$\alpha$ relation. The pioneering work of Lauer and
Postman (Lauer \& Postman 1992, 1994; Postman \& Lauer 1995) has demonstrated
the potential of BCGs in distance scale and peculiar velocity work. The detection
of very large-scale bulk streaming using BCGs has caused some to question
the global validity of the $L_m$-$\alpha$ relation (e.g., Riess, Press, \& Kirshner
1995b), but the verdict is not in yet. Ongoing work by Lauer and Postman,
in collaboration with Strauss, will greatly clarify the situation.
\end{enumerate}

I conclude by reiterating a point made at the outset of this Chapter.
The DIs discussed here are empirical relations whose physical
origins are only partially understood at best. There is a class of distance
indicators that are based on fairly rigorous physics, and whose absolute
calibration may be obtained from first principles. Gravitational lensing of
time-variable quasars and the Sunyaev-Zeldovich effect in clusters are
perhaps the most noteworthy of these. It is conceivable that these methods
will mature in the coming decade and add greatly to what we have learned
from the empirical DIs about the distance scale and the peculiar velocity
field. However, this additional information will most likely reinforce, 
rather than supplant, the knowledge obtained from the DIs
discussed here.

\bigskip
{\bf Acknowledgments:} I would like to thank David Burstein, Tod Lauer, 
Marc Postman,
Saul Perlmutter, and John Tonry for enlightening discussions
about the distance indicator relations in which they are
leading experts, and for providing me with data or postscript
for several of the figures presented here.

\vfill\eject
\def\aj{AJ}
\def\araa{ARA\&A}
\def\apj{ApJ}
\def\apjl{ApJ}
\def\apjs{ApJS}
\def\aap{A\&A}
\def\baas{BAAS}
\def\jrasc{JRASC}
\def\mnras{MNRAS}
\def\pasp{PASP}
\def\nat{Nature}
\def\iaucirc{IAU~Circ.}

\end{document}